\documentclass[prd,twocolumn,showpacs,preprintnumbers,nofootinbib,amsmath,amssymb,floatfix]{revtex4}

\usepackage{graphicx}% Include figure files
\usepackage{dcolumn}% Align table columns on decimal point
\usepackage{bm}% bold math
\usepackage{amsmath}

\newcommand{\beq}{\begin{eqnarray}}
\newcommand{\eeq}{\end{eqnarray}}

\newcommand{\real}{{\sf I}\kern-.12em{\sf R}}
\newcommand{\comp}{{\sf I}\kern-.50em{\sf C}}
\newcommand{\unity}{{\sf I}\kern-.54em{\sf 1}}

\def\spose#1{\hbox to 0pt{#1\hss}}
\def\ltapprox{\mathrel{\spose{\lower 3pt\hbox{$\mathchar"218$}}
 \raise 2.0pt\hbox{$\mathchar"13C$}}}

\begin{document}

\title{Analytic continuation of the critical line: suggestions for QCD}
\author{Paolo Cea}
\affiliation{
Dipartimento di Fisica dell'Universit\`a di Bari and INFN - Sezione di Bari, 
I-70126 Bari, Italy}
\email{paolo.cea@ba.infn.it}
\author{Leonardo Cosmai}
\affiliation{INFN - Sezione di Bari, I-70126 Bari, Italy}
\email{leonardo.cosmai@ba.infn.it}
\author{Massimo D'Elia, Chiara Manneschi}
\affiliation{Dipartimento di Fisica dell'Universit\`a di Genova 
and INFN - Sezione di Genova, I-16146 Genova, Italy}
\email{delia@ge.infn.it}
\author{Alessandro Papa}
\affiliation{Dipartimento di Fisica dell'Universit\`a della Calabria
and INFN - Gruppo collegato di Cosenza, 
I-87036 Arcavacata di Rende, Cosenza, Italy}
\email{papa@cs.infn.it}

\date{\today}% It is always \today, today,
             %  but any date may be explicitly specified

\begin{abstract}
We perform a numerical study of the systematic effects involved in the 
determination of the critical line at real baryonic chemical potential
by analytic continuation from results obtained at imaginary chemical 
potentials. We present results obtained in theories free of the sign problem,
such as two-color QCD with finite baryonic density and three-color QCD with 
finite isospin chemical potential, and comment on general features which 
could be relevant also to the continuation of the critical line in real QCD 
at finite baryonic density.
\end{abstract}


\maketitle

\section{Introduction}
\label{introd}

The study of QCD at nonzero baryonic density by numerical simulations
on a space-time lattice is plagued by the well-known sign problem:
the fermionic determinant is complex and the Monte Carlo sampling becomes
unfeasible.

One of the possibilities to circumvent this problem is to perform Monte
Carlo numerical simulations for imaginary values of the baryonic 
chemical potential, where the fermionic determinant is real and 
the sign problem is absent, and to infer the behavior at real chemical 
potential by analytic continuation.

The idea of formulating a theory at imaginary $\mu$ was first suggested 
in Ref.~\cite{Alford:1998sd}, while the effectiveness of the method
of analytic continuation was pushed forward in Ref.~\cite{Lombardo:1999cz}.
Since then, the method has been extensively applied to  
QCD~\cite{muim,immu_dl,azcoiti,chen,defor06,Wu:2006su,sqgp,2im} and tested 
in QCD-like theories free of the sign problem~\cite{Hart:2000ef,giudice,
cea,cea1,conradi,Shinno:2009jw} and in spin models~\cite{potts3d,kt}. 

The state-of-the-art is the following: 
\begin{enumerate}
\item[(i)] the method is well founded and works fine within the limitations posed 
by the presence of nonanalyticities and by the periodicity of 
the theory with imaginary chemical potential~\cite{rw}; 
\item[(ii)] the analytic continuation of physical observables is improved if ratios 
of polynomials (or Pad\'e approximants~\cite{Lombardo:2005ks}) are used as 
interpolating functions at imaginary chemical potential~\cite{cea,cea1}; 
\item[(iii)] the analytic continuation of the (pseudo-)critical line on the
temperature -- chemical potential plane is well justified, but a careful
test in two-color QCD~\cite{cea1} has cast some doubts on its reliability.
\end{enumerate}

In particular, the numerical analysis in two-color QCD of Ref.~\cite{cea1} 
has shown that, while there is no doubt that an analytic function exists which 
interpolates numerical data for the pseudocritical couplings for both
imaginary and real $\mu$ across $\mu=0$, determining this function
by an interpolation of data at imaginary $\mu$ could be misleading.
Indeed, in the case of polynomial interpolations, there is a clear indication
in two-color QCD that nonlinear terms in $\mu^2$ play a relevant role at 
real $\mu$, but are less visible at imaginary $\mu$, thus calling for an 
accurate knowledge of the critical line there and, consequently, for
very precise numerical data.

The above described scenario could well be peculiar to two-color QCD 
and strongly depends on the choice of parameters of Ref.~\cite{cea1}. Therefore,
in this work we perform a systematic study of the analytic continuation of 
the critical line 
\begin{enumerate}
\item[(i)] in two-color QCD with a fermionic mass different than in 
Ref.~\cite{cea1},
\item[(ii)] in another sign-free theory, SU(3) with a nonzero density of isospin.
\end{enumerate}  
The aim of this study is to single out some general features of the 
analytic continuation of the critical line and to understand if
and to what extent they can apply also to the physically relevant case of QCD.

\begin{figure}[htbp]
\includegraphics*[height=0.295\textheight,width=0.95\columnwidth]
{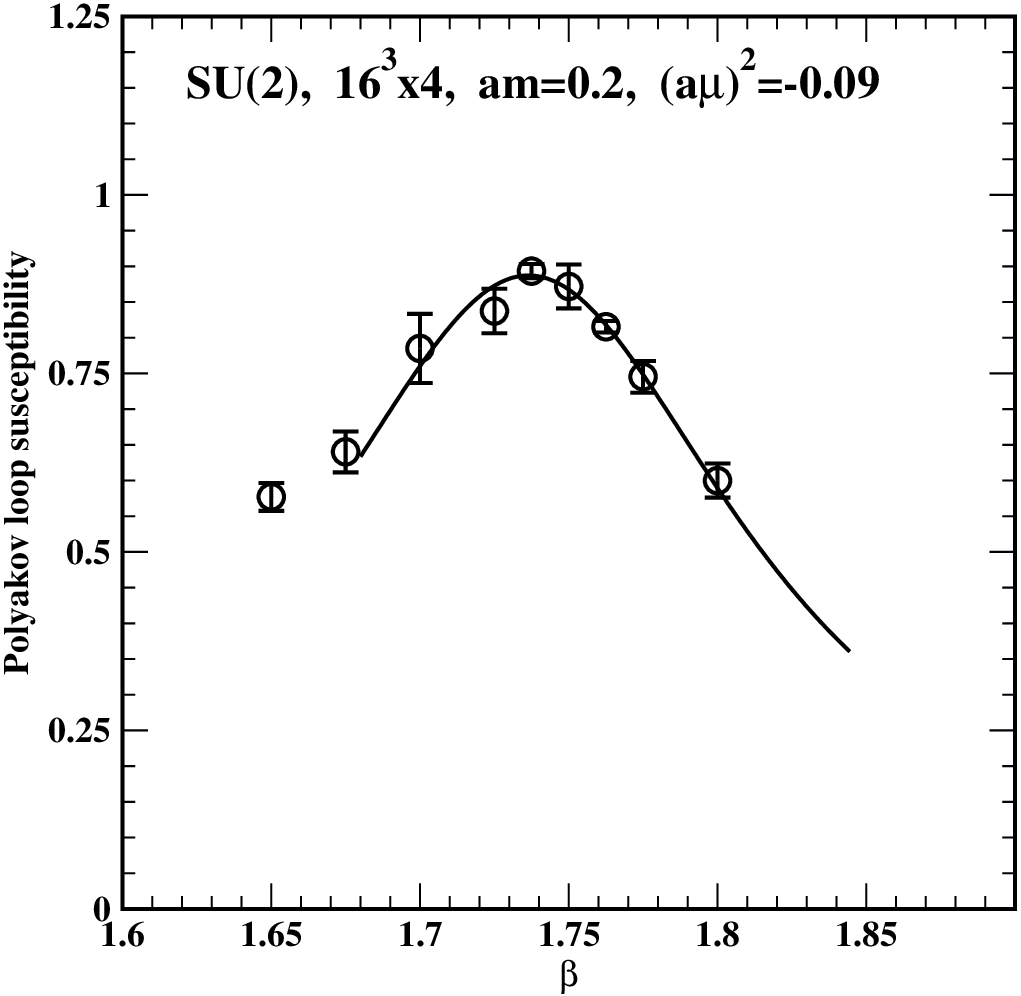}
\includegraphics*[height=0.295\textheight,width=0.95\columnwidth]
{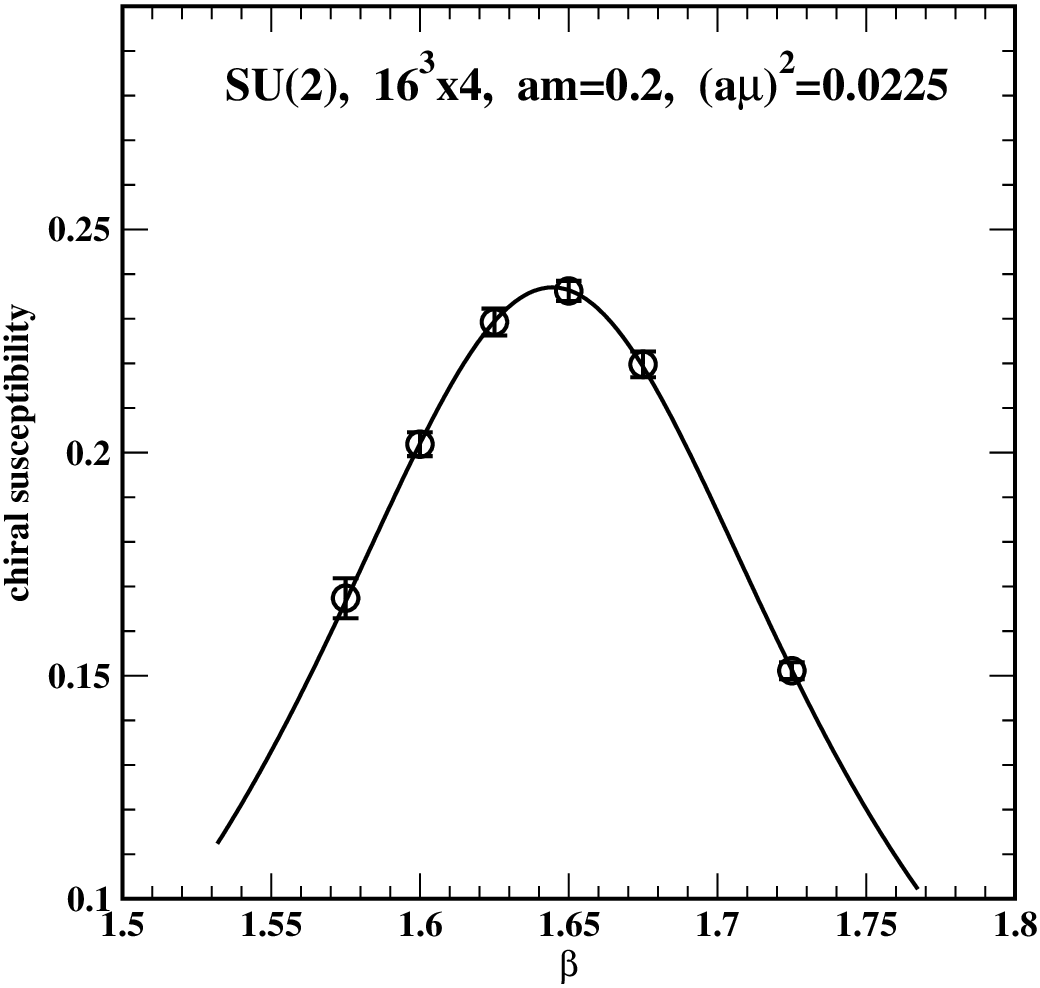}
\includegraphics*[height=0.295\textheight,width=0.95\columnwidth]
{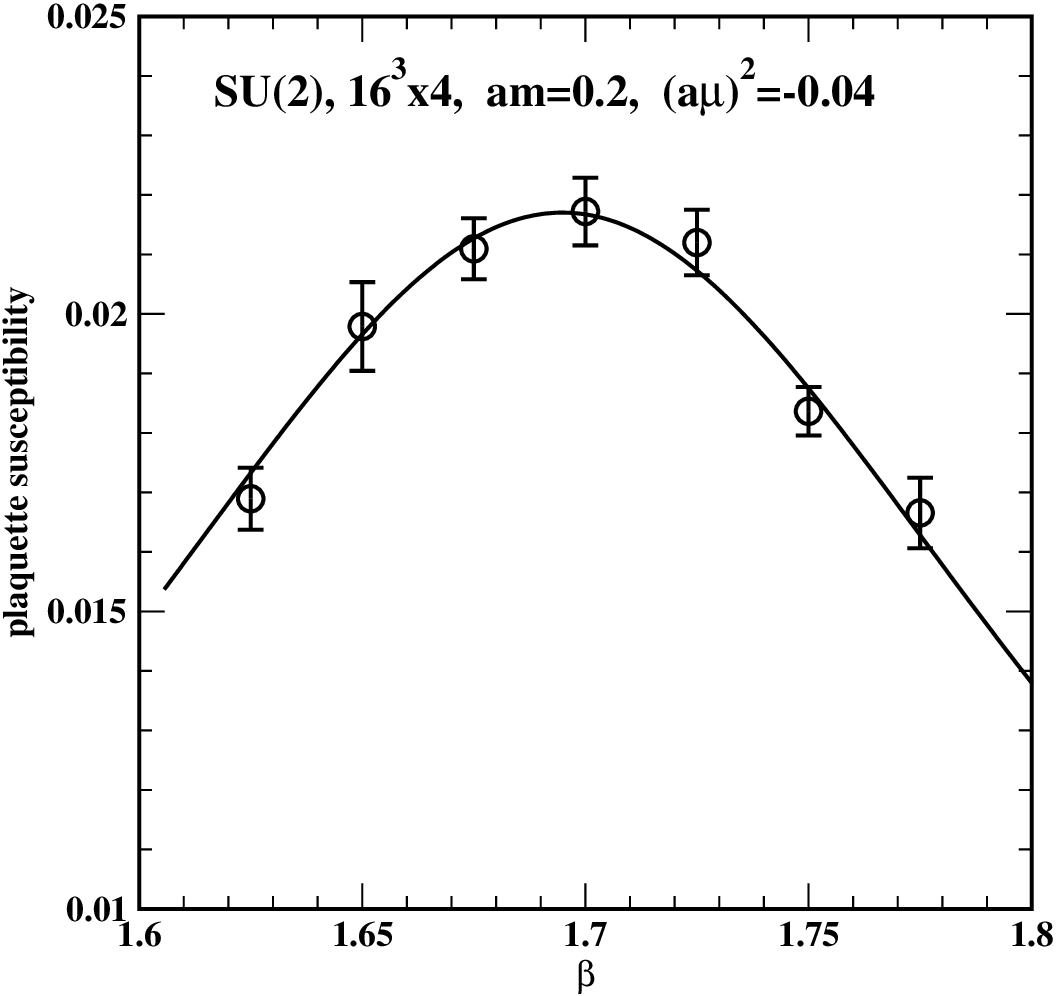}
\caption{Susceptibility of Polyakov loop, chiral condensate, plaquette
{\it vs} $\beta$ in SU(2) on a 16$^3\times 4$ lattice with $am$=0.2
and $(a\mu)^2 = -0.09$ (for the Polyakov loop), $(a\mu)^2 = 0.0225$ (for
the chiral condensate) and $(a\mu)^2 = -0.04$ (for the plaquette). The solid
lines represent the Lorentzian interpolation.}
\label{su2_nt=4_peak}
\end{figure}

\section{Analytic continuation of the critical line in two-color QCD}
\label{su2}

As in Ref.~\cite{cea}, to which we refer for further details about our
numerical setup, we have performed Monte Carlo simulations of the SU(2) gauge 
theory with $N_f=8$ degenerate staggered fermions. The algorithm adopted 
has been the usual exact $\phi$ algorithm described in 
Ref.~\cite{Gottlieb:1987mq}, properly modified for the inclusion of a finite 
chemical potential. 

The general strategy is the following: first the critical line is
sampled by determining, for a set of $\mu^2$ values, the critical couplings 
$\beta_c (\mu^2)$ for which the susceptibilities of a bulk observable exhibit 
a peak. Since the location of the critical line may well 
depend on the chosen observable, we have used three probe observables: the 
chiral condensate, the Polyakov loop and the plaquette.
The $\mu^2<0$ values are chosen to lie inside the so-called first 
Roberge-Weiss (RW) sector, i.e. in the range delimited by 
$(a\mu)^2=-(\pi/8)^2$ and $\mu^2=0$ (see Ref.~\cite{cea} for a detailed 
discussion on the phase diagram in the $T - i \mu$ plane).

Then, the critical line is guessed by interpolating the values of 
$\beta_c(\mu^2)$ for $\mu^2 \leq 0$ only. The validity of the interpolation
is evaluated by comparing its analytic continuation to the region 
$\mu^2 > 0$ with the direct determinations of the critical coupling in this 
region.

The procedure above is repeated for each of the three probe observables 
considered.

\subsection{$N_t = 4$, $am = 0.07$}

The SU(2) gauge theory with $N_f=8$ degenerate staggered fermions of mass 
$am=0.07$ has been studied on a $16^3\times 4$ lattice in Ref.~\cite{cea1}.
We found that
\begin{enumerate}
\item[(i)] there is a very weak dependence of the values of $\beta_c(\mu^2)$ 
on the probe observable;
\item[(ii)] there is no better interpolation of $\beta_c(\mu^2)$ data at
$\mu^2\leq 0$ than a polynomial of the form $a_0+a_1(a\mu)^2$;
\item[(iii)] the extrapolation to the region $\mu^2 > 0$ overshoots the direct
determinations of $\beta_c(\mu^2)$ in that region, the discrepancy becoming
larger and larger for increasing $\mu^2$;
\item[(iv)] data for $\beta_c(\mu^2)$ for {\em both} $\mu^2\leq 0$ and $\mu^2>0$ 
can be nicely interpolated by an analytic function (a polynomial 
of third order in $\mu^2$  works nicely), this excludes the option 
that the critical line be not smooth and possibly nonanalytic in the
thermodynamic limit.
\end{enumerate}

\begin{table*}[htbp]
\setlength{\tabcolsep}{0.5pc}
\centering
\caption[]{Summary of the values of $\beta_c(\mu^2)$ obtained by fitting the 
peaks of the susceptibilities of chiral condensate 
$\langle\overline\psi\psi\rangle$, Polyakov loop $\langle L \rangle$ and 
plaquette $\langle P \rangle$ in SU(2) on a 16$^3\times 4$ lattice with
fermionic mass $am$=0.2. For each interpolation the $\chi^2/{\rm d.o.f.}$
is given.}
\begin{tabular}{ddcdcdc}
\hline
\hline
\multicolumn{1}{c}{\hspace{0.70cm}$(a\mu)^2$} &
\multicolumn{1}{c}{\hspace{1cm}$\langle\overline\psi\psi\rangle$} &
$\chi^2/{\rm d.o.f.}$ &
\multicolumn{1}{c}{\hspace{1cm}$\langle L \rangle$} &
$\chi^2/{\rm d.o.f.}$ &
\multicolumn{1}{c}{\hspace{1cm}$\langle P \rangle$} &
$\chi^2/{\rm d.o.f.}$ \\
\hline
-0.1452 & 1.8098(71) & 1.01 & 1.7991(14) & 0.51 & 1.8058(67) & 1.13 \\
-0.1225 & 1.7713(30) & 0.97 & 1.7709(30) & 0.21 & 1.7688(43) & 0.20 \\
-0.09   & 1.7465(37) & 0.30 & 1.7365(76) & 0.66 & 1.7448(60) & 0.79 \\
-0.04   & 1.6998(42) & 0.92 & 1.6934(53) & 2.10 & 1.6948(86) & 0.73 \\
-0.01   & 1.6762(41) & 0.37 & 1.6680(86) & 0.20 & 1.6745(77) & 0.24 \\
 0.     & 1.6694(44) & 0.30 & 1.6634(76) & 1.50 & 1.6709(71) & 0.58 \\
 0.0225 & 1.6445(20) & 0.03 & 1.6180(65) & 1.01 & 1.6440(34) & 0.14 \\
 0.04   & 1.6430(54) & 1.04 & 1.6105(48) & 1.87 & 1.6367(67) & 0.74 \\
 0.09   & 1.5989(45) & 0.30 & 1.5606(70) & 0.73 & 1.6005(81) & 0.68 \\
\hline
\hline
\end{tabular}
\label{su2_nt=4_data}
\end{table*}

\begin{table}[htbp]
\setlength{\tabcolsep}{0.25pc}
\centering
\caption[]{Parameters of the fits to the critical 
couplings in SU(2) on a 16$^3\times 4$ lattice with fermionic mass $am$=0.2
according to the fit function $\beta_c (\mu^2) = a_0 + a_1 (a\mu)^2$, for each
of the three probe observables used (chiral condensate, Polyakov loop
and plaquette).}
\begin{tabular}{cddc}
\hline
\hline
observable & \multicolumn{1}{c}{$a_0$} & \multicolumn{1}{c}{$a_1$} 
& $\chi^2$/d.o.f. \\
\hline
$\langle\overline\psi\psi\rangle$ & 1.6671(28) & -0.876(32) & 1.61 \\
$\langle L \rangle$               & 1.6556(44) & -0.981(33) & 1.34 \\
$\langle P \rangle$               & 1.6656(48) & -0.885(49) & 1.40 \\
\hline
\hline
\end{tabular}
\label{su2_polin2_fit}
\end{table}

The conclusion we drew in Ref.~\cite{cea1} was that terms of order $\mu^4$ 
and $\mu^6$ play a relevant role at $\mu^2 > 0$, but are less visible at 
$\mu^2 < 0$, thus calling for an accurate knowledge of the critical line in 
all the first RW sector.

Verifying if this scenario is peculiar to SU(2) with the parameters choice 
as in Ref.~\cite{cea1} is the main motivation of the present work.
For this reason, in the following subsection we repeat the same analysis 
of Ref.~\cite{cea1} in SU(2) with a different numerical set up, whereas 
in the next section we turn to a completely different theory.

\begin{figure}[htbp]
\includegraphics*[height=0.295\textheight,width=0.95\columnwidth]
{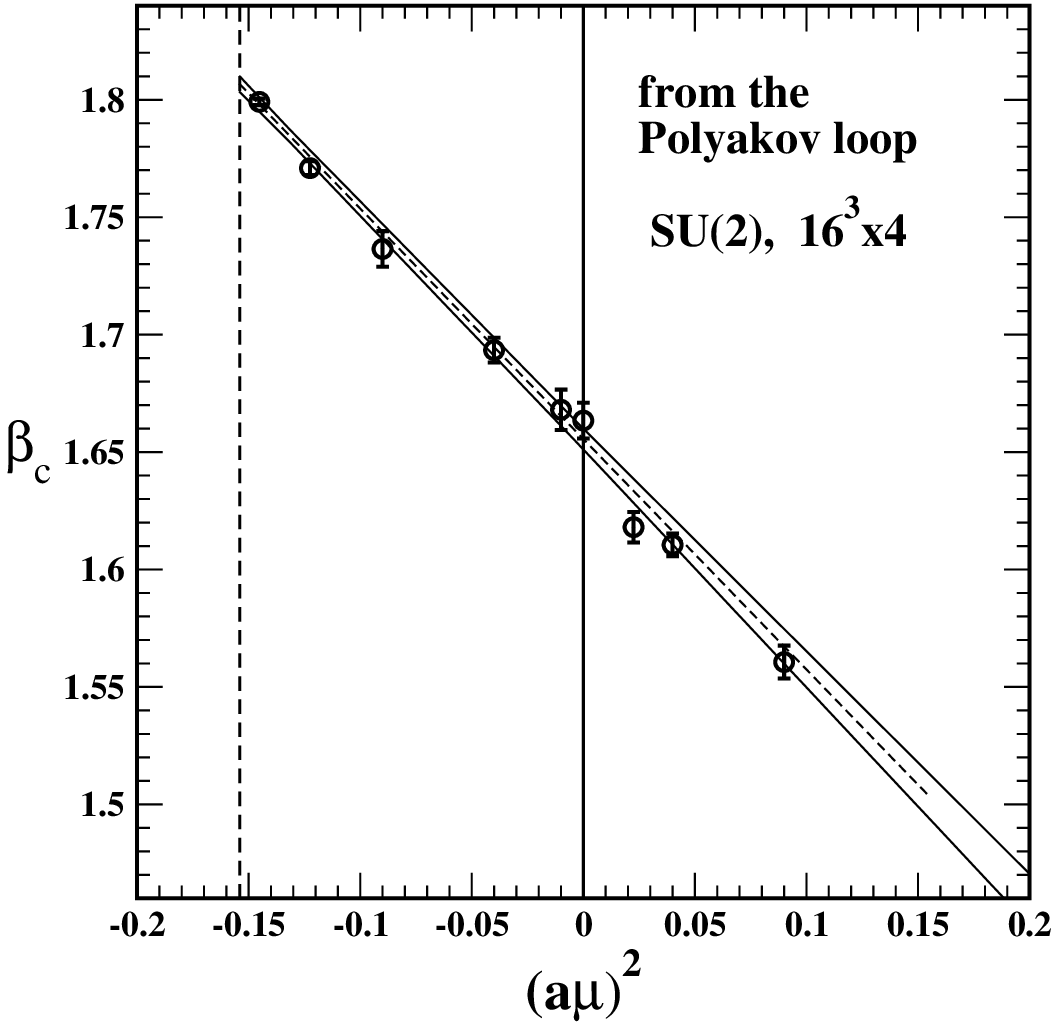}
\includegraphics*[height=0.295\textheight,width=0.95\columnwidth]
{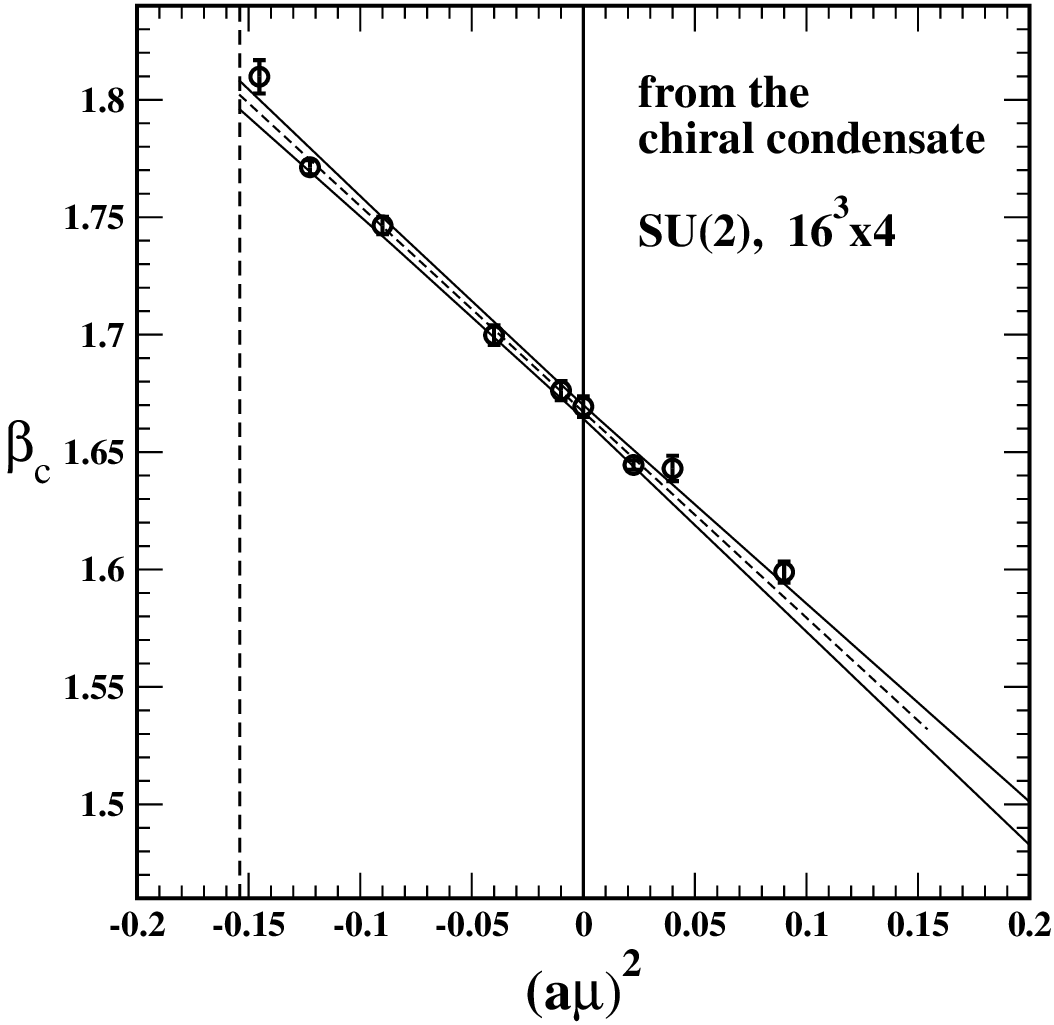}
\includegraphics*[height=0.295\textheight,width=0.95\columnwidth]
{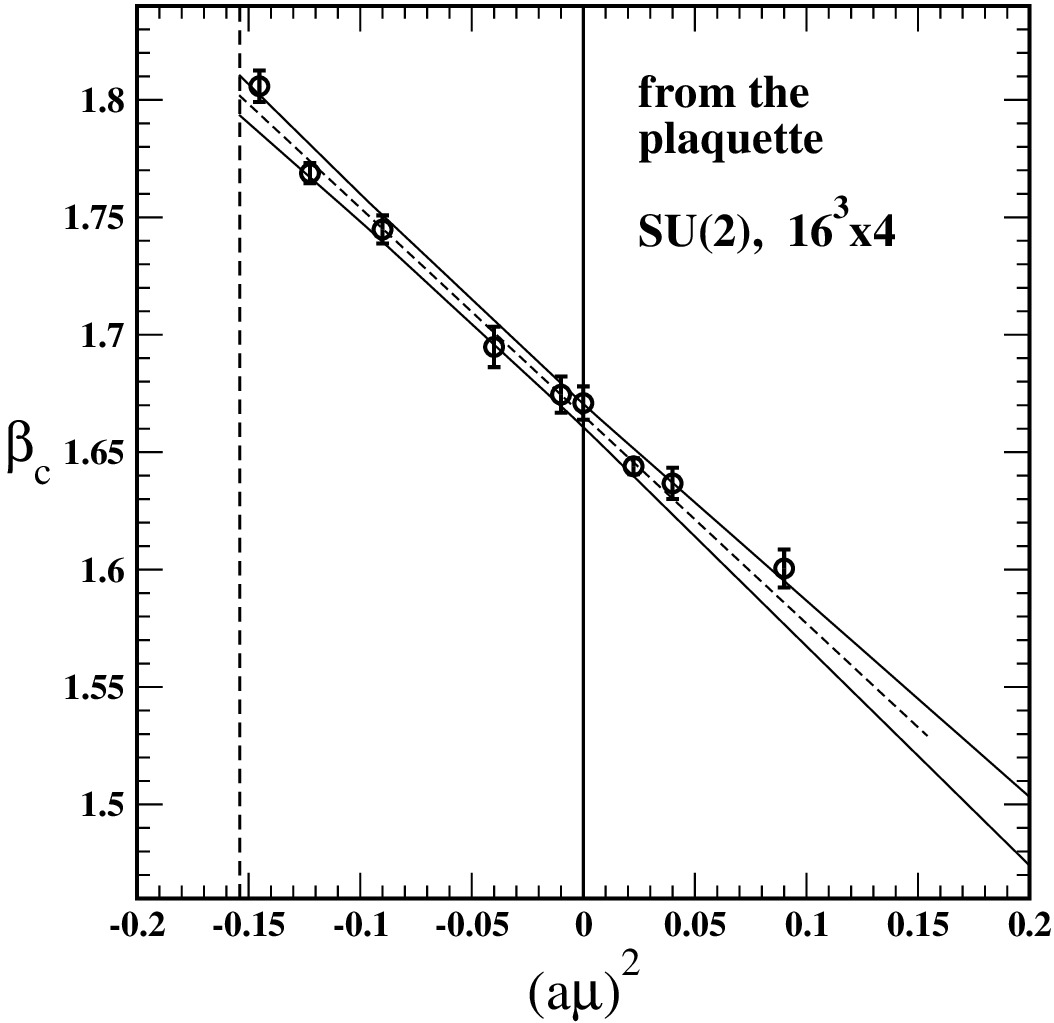}
\caption{Critical couplings obtained from the susceptibilities of
Polyakov loop, chiral condensate and plaquette in SU(2) on a 
16$^3\times 4$ lattice with $am$=0.2, together
with a linear fit (dotted line) in $(a\mu)^2$ to data with $\mu^2 \leq 0$.
Here and in the following plots of the same kind, the solid lines around the 
best fit line delimit the 95\% CL region.}
\label{su2_nt=4_crit}
\end{figure}

\begin{figure*}[htbp]
\includegraphics*[height=0.295\textheight,width=0.95\columnwidth]
{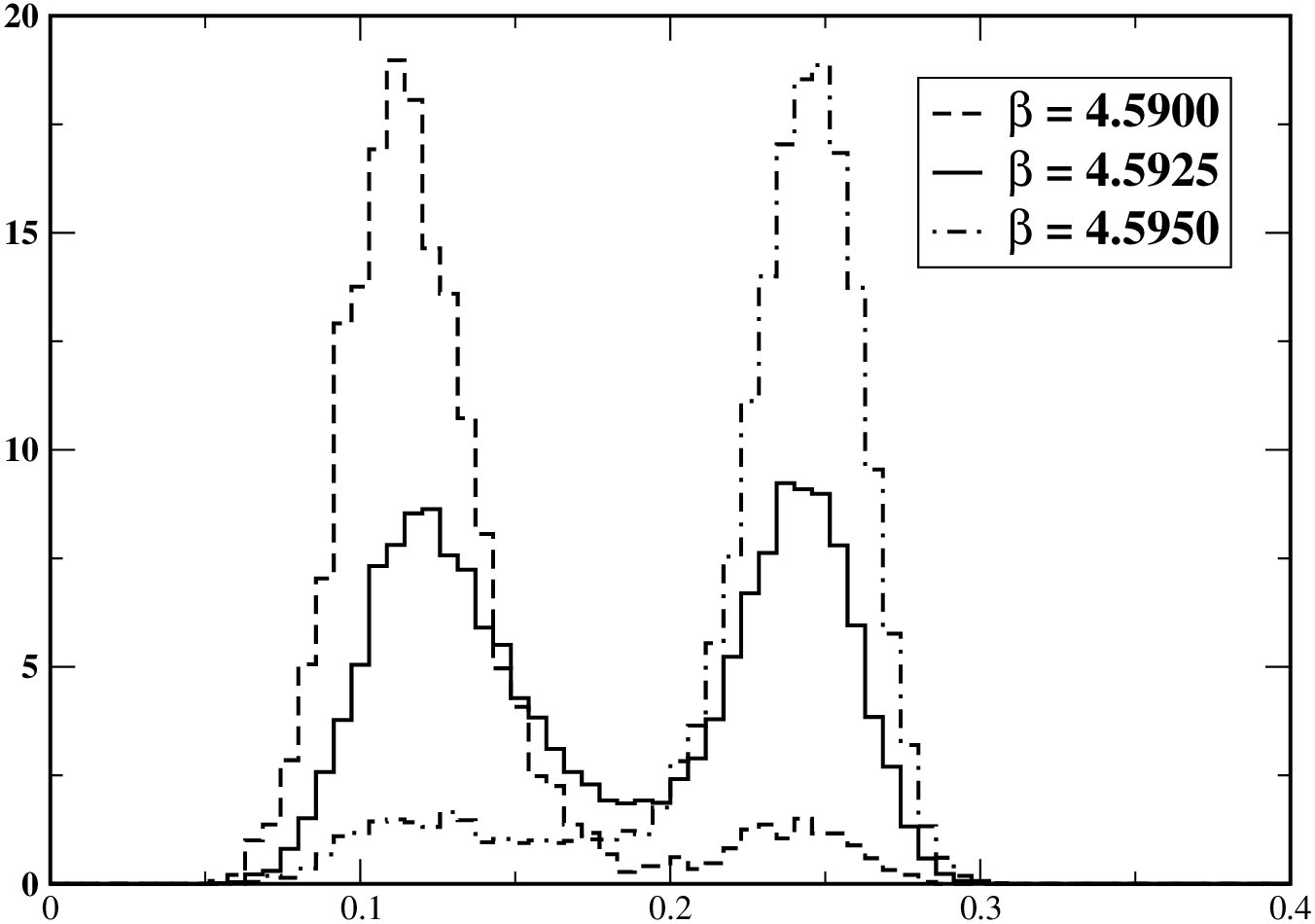}
\includegraphics*[height=0.295\textheight,width=0.95\columnwidth]
{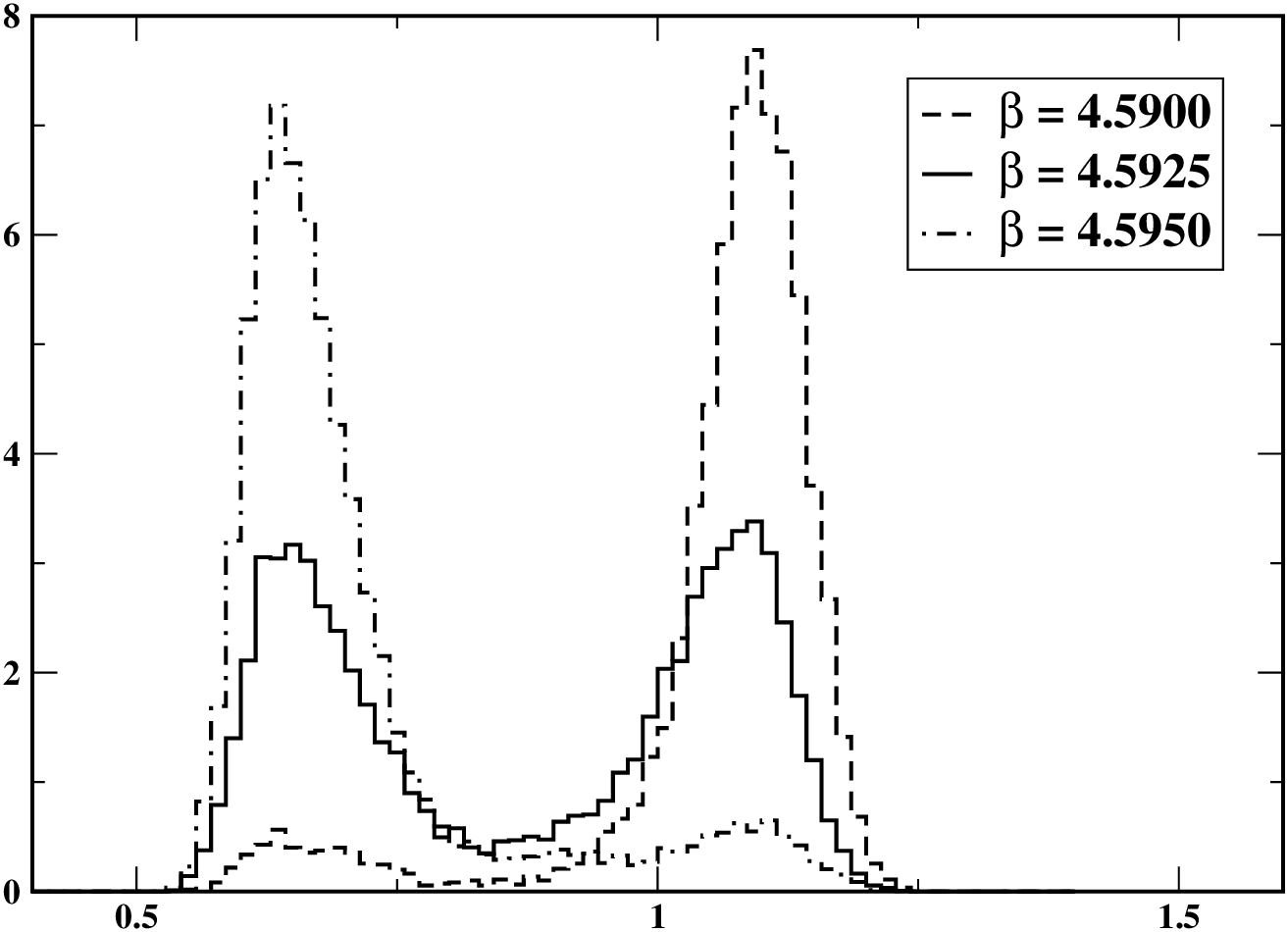}
\caption{Distributions of the real part of the Polyakov loop (left) and of
the chiral condensate (right) in SU(3) with finite isospin density on 
a 8$^3\times 4$ lattice with $am$=0.1 at $\mu^2/(\pi T)^2$=0.16 and for 
three $\beta$ values around the transition.}
\label{su3_distributions}
\end{figure*}

\begin{figure}[htbp]
\includegraphics*[height=0.295\textheight,width=0.95\columnwidth]
%{./figures/su3_poly_history.eps}
{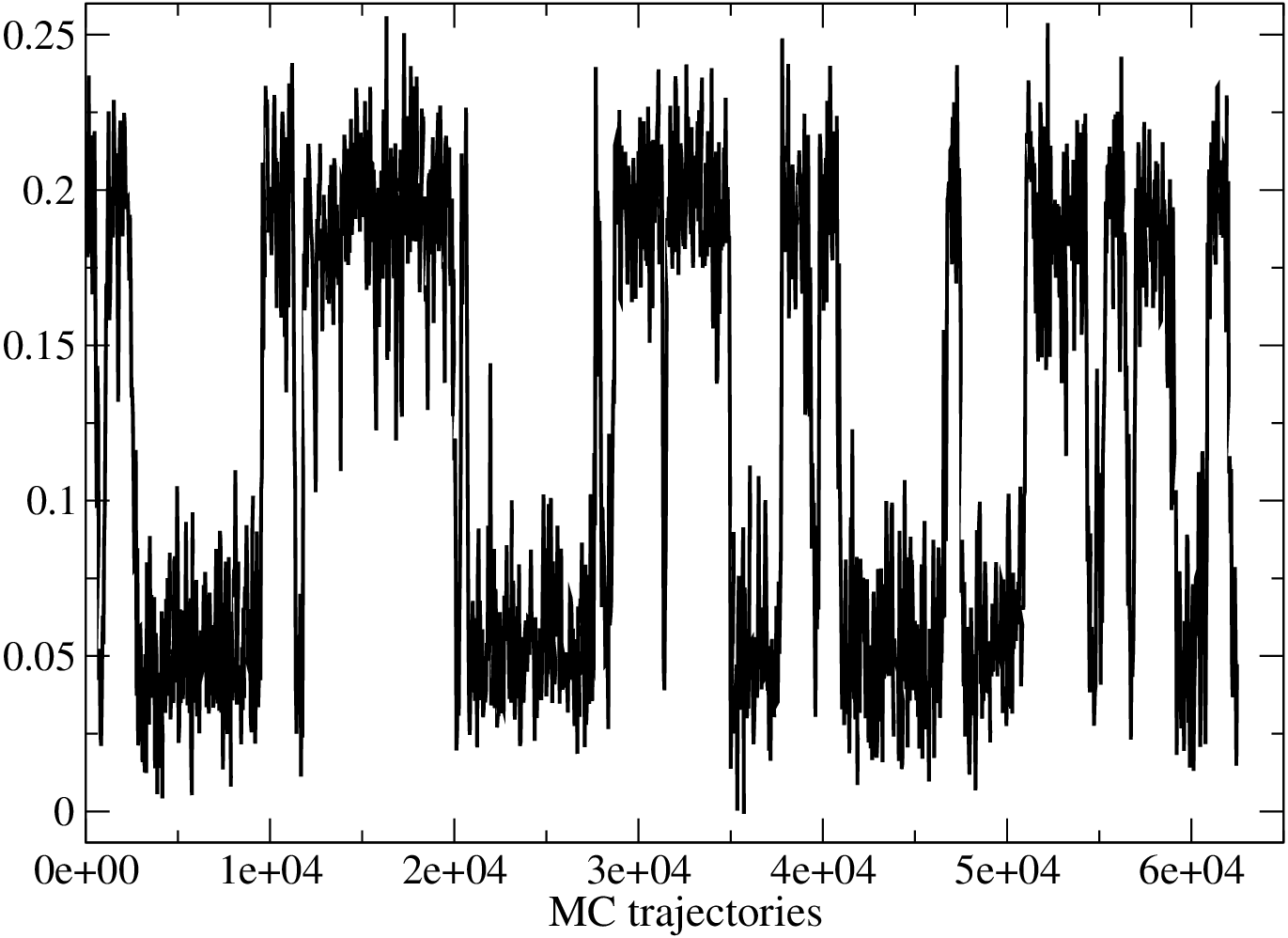}
\caption{History of the Monte Carlo run of the real part of the Polyakov loop 
in SU(3) with finite isospin density on a 8$^3\times 4$ lattice with 
$am$=0.1 at $\mu^2/(\pi T)^2=-0.09$ and $\beta$=4.750.}
\label{su3_poly_history}
\end{figure}

\subsection{$N_t = 4$, $am = 0.2$}

We have considered SU(2) with $N_f=8$ degenerate staggered fermions of
mass $am=0.2$ on a $16^3\times 4$ lattice. The reason for setting a larger
value for the quark mass is simple: at any fixed temperature $T$ the
critical value of $\mu$ gets larger, so that the critical line could become
less curved in the physical region $\mu^2>0$. This should reduce the
importance of higher orders in $\mu^2$ in the description of the critical
line by a polynomial.

As in Ref.~\cite{cea1}, the critical line is sampled by looking for 
peaks in the susceptibilities of Polyakov loop, chiral condensate and
plaquette. In Fig.~\ref{su2_nt=4_peak} we show as example the behavior 
in $\beta$ of the susceptibility of the three observables at different 
values of $\mu^2$, together with the Lorentzian interpolation of the peak. 
The statistics for each data point is about 10-20k trajectories
of one Molecular Dynamics (MD) time length.

All determinations of $\beta_c(\mu^2)$ are summarized in 
Table~\ref{su2_nt=4_data}, from which we see a mild dependence
on the probe observable for $\mu^2\leq 0$; at $\mu^2>0$, instead, 
determinations from the Polyakov loop are systematically below those 
from the chiral condensate and the plaquette, which instead agree within 
errors.

As in Ref.~\cite{cea1}, the best interpolation of $\beta_c(\mu^2)$ data at
$\mu^2\leq 0$ is a polynomial of the form $a_0+a_1(a\mu)^2$. In 
Table~\ref{su2_polin2_fit} we report the values of the fit parameters 
and their uncertainties as obtained with the MINUIT minimization code.
The extrapolation to $\mu^2> 0$ compares very well with the direct 
determinations of $\beta_c(\mu^2)$ in that region, as shown in 
Figs.~\ref{su2_nt=4_crit}. In these figures, and in the following of the
same kind, the band around the best fit (dotted line) represents the 
95\% confidence level (CL) region.

We can draw some partial conclusions about the analytic continuation of the 
critical line in SU(2):
\begin{itemize}
\item for our given accuracy, there seems to be no room for quartic and 
sextic terms in $\mu^2$ in the interpolation of $\beta_c(\mu^2)$ data for 
imaginary $\mu$;
\item the extrapolation to $\mu^2>0$ works definitely better for larger 
masses, i.e. away from the chiral limit.
\end{itemize}

Both these features could be peculiar to the SU(2) theory. This motivates
the need to repeat a similar analysis in a different theory, sharing with
SU(2) the property of being free of the sign problem.  

\section{Analytic continuation of the critical line in 
three-color QCD at finite isospin chemical potential}

The SU(3) gauge theory with a finite density of isospin charge~\cite{ks}
is a theory in which 
the chemical potential is $\mu_{\rm iso}$ for half of the fermionic species 
and $-\mu_{\rm iso}$ for the other half. The partition function, which
is even in $\mu_{\rm iso}$ and depends only on $\mu_{\rm iso}^2$, can be 
written as follows:
\beq
Z(T,\mu_{\rm iso}) = \int \mathcal{D}U e^{-S_{G}} 
\det M [\mu_{\rm iso}]
\det M [-\mu_{\rm iso}]
\label{partfun}
\eeq
where the integration is over gauge link variables,
$S_G$ is the pure gauge action and $M$ the fermion matrix
(we adopt a standard staggered discretization).
This leads to a real and positive measure, because of the 
property $\det M [-\mu_{\rm iso}] = (\det M [\mu_{\rm iso}])^*$,
and therefore to a theory free of the sign problem. This theory is 
obviously closer to real QCD than two-color QCD, being yet unphysical,
since it implies a zero baryonic density, while in nature a nonzero
isospin density is always accompanied by a nonzero 
%(and much larger)
baryonic density; moreover the isospin charge is not a conserved
number in the real world.

Nevertheless, for our purposes this theory is very convenient since it
provides us with another theoretical laboratory for the method of analytic 
continuation.

Similarly to SU(2) with finite baryonic density, at imaginary values of the
chemical potential $\mu_{\rm iso}$ (which we will call simply $\mu$ from now 
on) the theory exhibits RW-like transition lines (we refer to the following 
subsection for a discussion about the determination of their position). 
The first RW sector is given by the strip $- (0.5)^2 \lesssim \mu^2/(\pi T)^2 
\leq 0$.

\begin{figure}[htbp]
\includegraphics*[height=0.295\textheight,width=0.95\columnwidth]
{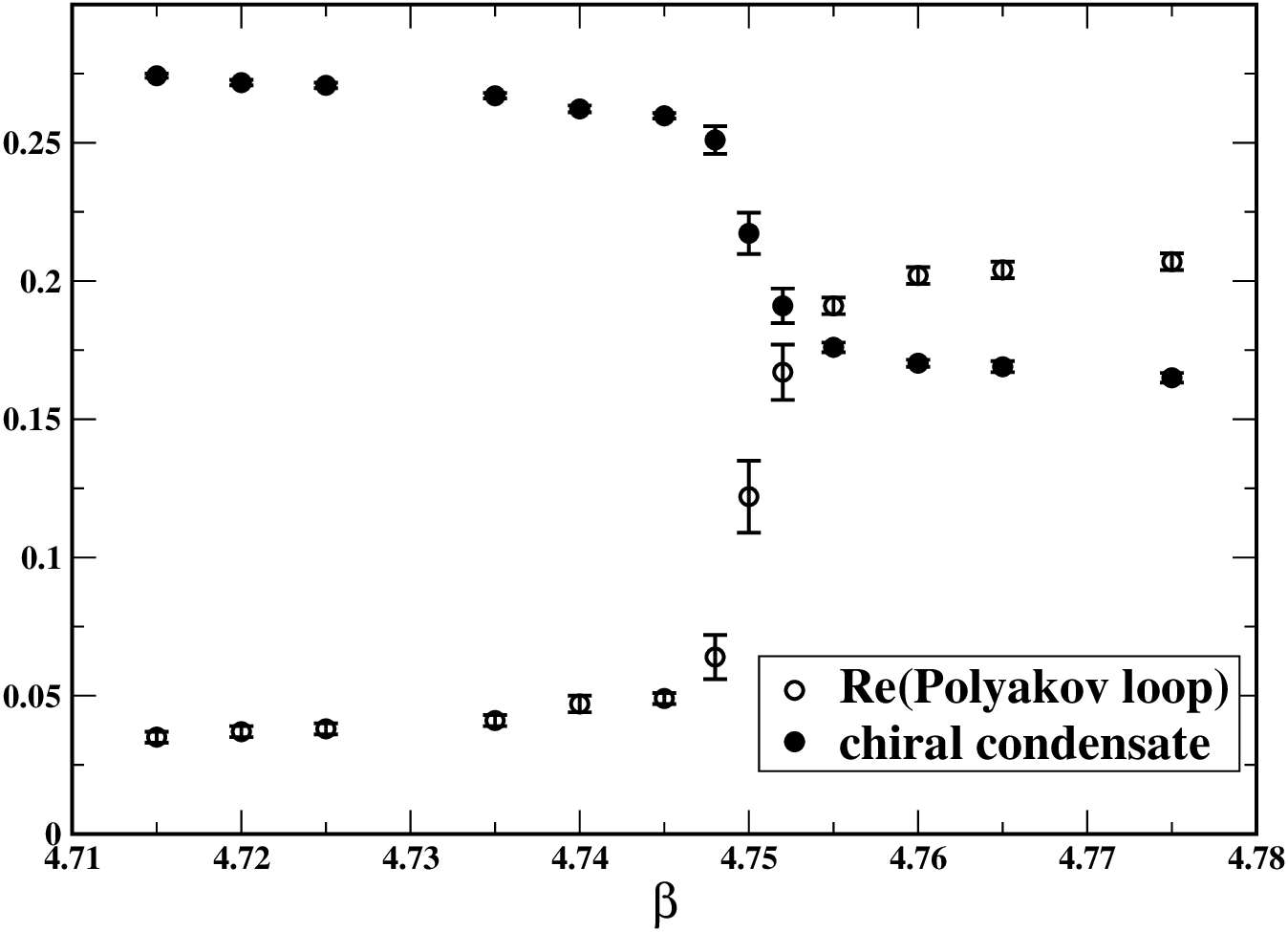}
\caption{Real part of the Polyakov loop and chiral condensate {\it vs} 
$\beta$ in SU(3) with finite isospin density at $\mu^2/(\pi T)^2=-0.09$ on a 
8$^3\times 4$ lattice with $am$=0.1.}
\label{su3_transition}
\end{figure}

In our numerical analysis, we consider finite isospin SU(3) with $N_f=8$ 
degenerate staggered fermions of mass $am=0.1$ on a $8^3\times 4$ lattice. 
Differently from SU(2) with finite baryonic density, the critical line
in the temperature -- chemical potential plane is a line of (strong)
first order transitions, over all the investigated range of $\mu^2$ values,
$-0.2304 \leq \mu^2/(\pi T)^2 \leq 0.2025$. This is one of the reasons 
for working on a smaller volume than in the SU(2) case (tunneling
between the different phases would have been sampled with much more 
difficulty on a larger volume) and clearly emerges from the
distribution on the thermal equilibrium ensemble of the values of
observables like the (real part of) the Polyakov loop, the chiral condensate, and
the plaquette across the transition (see Fig.~\ref{su3_distributions}
as an example showing the typical two-peak structures). As a further
evidence of the first order nature of the transition, we show 
in Fig.~\ref{su3_poly_history} the Monte Carlo run history of the 
(real part of) the Polyakov loop at $\mu^2/(\pi T)^2 =-0.09$ and 
$\beta=\beta_c=4.750$, which exhibits tunneling events between 
the two phases every few thousands trajectories, on average. 
Typical statistics have been around 10K trajectories of 1 MD unit
for each run, growing up to 100K trajectories for 2-3 $\beta$ values
around $\beta_c(\mu^2)$, for each $\mu^2$, in order to correctly sample
the critical behavior at the transition.
The critical
$\beta(\mu^2)$ is determined as the point where the two peaks 
have equal height and, in all the cases considered, this point turned out
to be the same for all the adopted observables. In particular
this means that, at the transition, both the chiral condensate 
and the (real part of) the Polyakov loop ``jump'' at the same point
(see Fig.~\ref{su3_transition}, as an example).

If one looks at the ``gap'' between the two peaks at the transition (determined
as the difference between the values of an observable at the maxima 
of the right and left distribution peaks), it turns out that the dependence of this gap on 
$\mu^2/(\pi T)^2$ in the considered range is slightly observable-dependent,
being roughly constant for the Polyakov loop and increasing for the
chiral condensate and the spatial plaquette, up to a turning point at
$\mu^2/(\pi T)^2 \simeq 0.1$, where the strength of the transition
seems to reach a maximum. In particular, no critical endpoint
for the first order line is detectable, within the explored range. 
We notice that in Ref.~\cite{wenger}, where the same theory 
has been explored even if with a slightly larger quark mass
($am = 0.14$), a critical endpoint was reported, but for a real
chemical isospin chemical potential ($\mu/T \sim 2.5$) which is well outside
the range explored by us.

\begin{figure}[htbp]
\includegraphics*[height=0.295\textheight,width=0.95\columnwidth]
{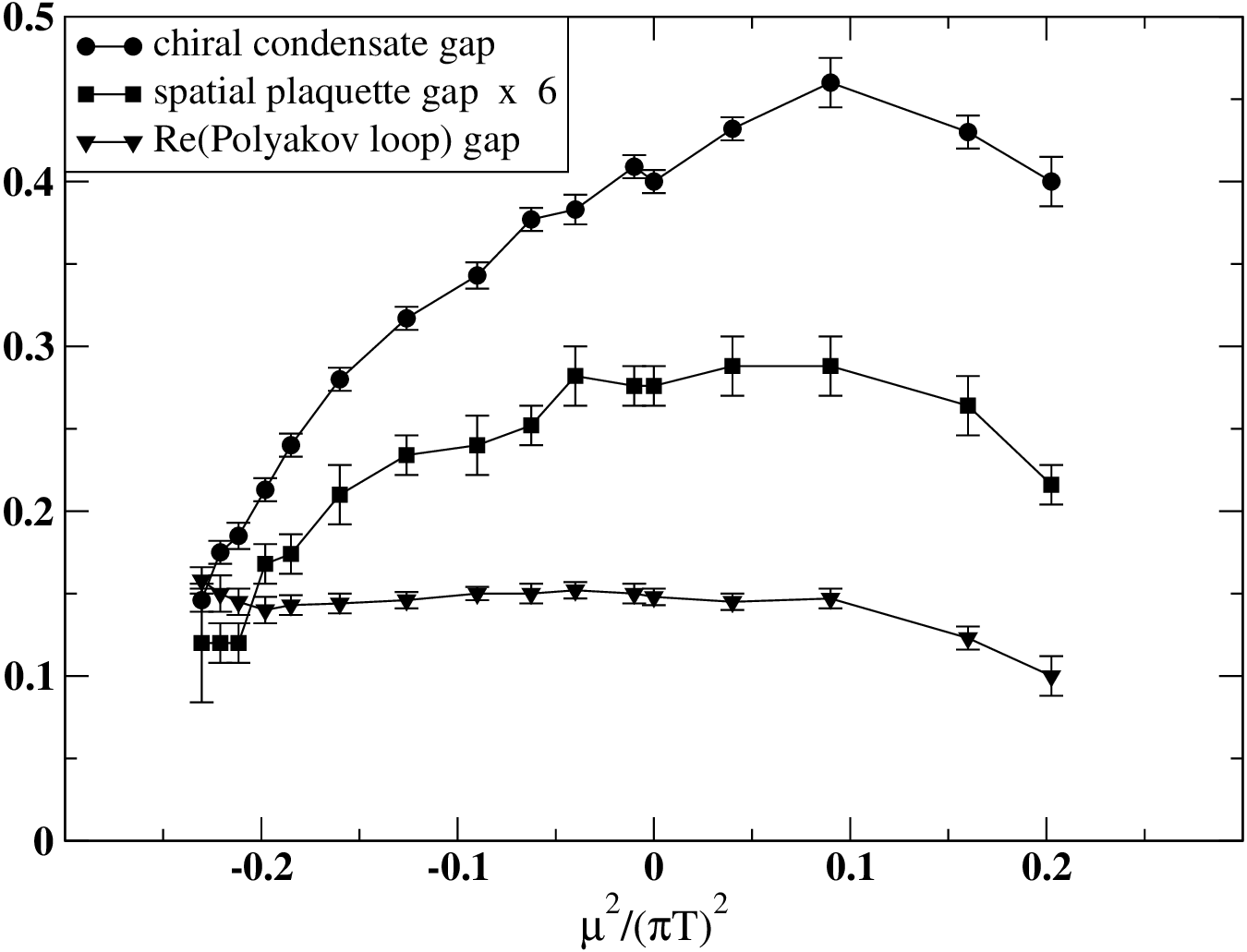}
\caption{Gap of several observables at the transition for several values of 
the isospin chemical potential.}
\label{su3_gap}
\end{figure}

\begin{figure}[htbp]
\includegraphics*[height=0.295\textheight,width=0.95\columnwidth]
{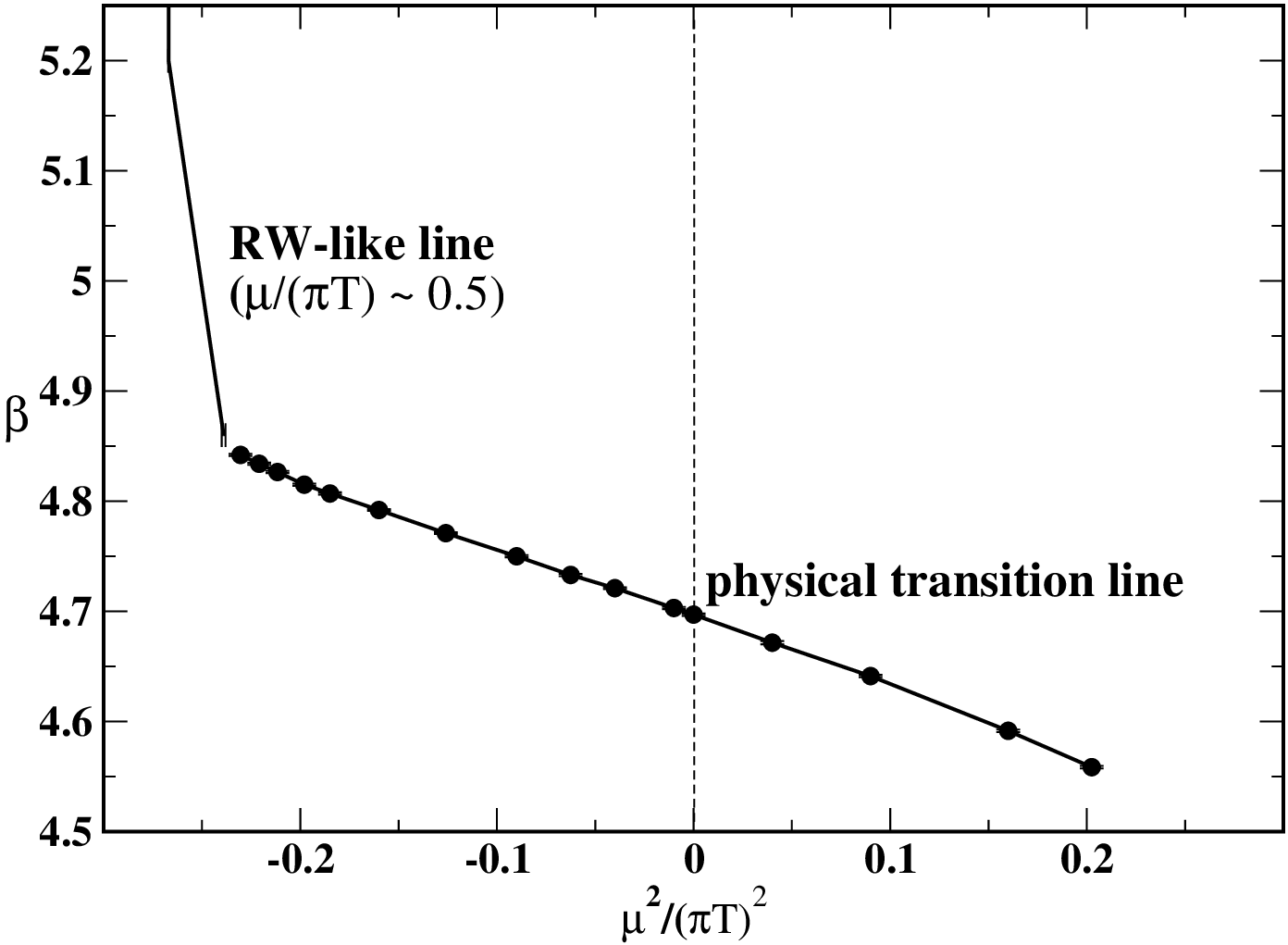}
\caption{Sketch of the phase diagram of SU(3) with finite isospin density 
on a 8$^3\times 4$ lattice with $am$=0.1.}
\label{su3_phase_diag}
\end{figure}

When combining the first RW-like transition line with the numerical
determinations 
for the critical $\beta$'s (see Sec.~\ref{su3_results}), the
resulting phase diagram in the ($\beta,\mu^2/(\pi T)^2$) looks like
in Fig.~\ref{su3_phase_diag}. 

\subsection{RW-like transitions at finite isospin density}

In order to understand the unphysical, RW-like phase transition taking place
in the high temperature region of the QCD phase diagram 
in the presence of an imaginary
isospin chemical potential, let us first remember the origin of the
usual RW transition in the presence of an imaginary baryon chemical
potential. That is linked to the dynamics of the Polyakov loop.
Let us define $\theta \equiv {\rm Im}(\mu)/T$, where $\mu$ is the
baryon chemical potential, and 
let us call $e^{\phi_j}$, with $j = 1, \dots N_c$, the eigenvalues of
the untraced Polyakov loop, which satisfy the condition 
$\sum_j \phi_j = 0_{\rm mod 2\pi}$.
The one-loop effective potential of the untraced Polyakov loop in presence 
of the baryon chemical potential is then given by~\cite{rw}
\begin{eqnarray}
&& V_{\rm eff} = \frac{\pi^2T^4}{24}\sum_{j,k=1}^{N_c}
\left(1-\left(\left[\frac{\phi_{j}}{\pi}-\frac{\phi_{k}}{\pi}\right]_{\rm mod2}
-1\right)^{2}\right)^{2}\nonumber\\
&& - \frac{\pi^2T^4}{12} \sum_{i=1}^{N_f} \sum_{j=1}^{N_c}
\left(1-\left(\left[\frac{\phi_{j}
+ \theta}{\pi}+1\right]_{\rm mod2}-1\right)^{2}\right)^{2} 
\label{effpotbar}
\end{eqnarray}

The first term is the gluonic contribution, which has three degenerate
minima, corresponding to SU($N_c$) matrices belonging to the center
of the group. In absence of fermions that means spontaneous breaking
of center symmetry. For $\theta = 0$ the fermion contribution is
minimum for $\phi_j = 0 \,\, \forall j$, and the center element
corresponding to a real Polyakov loop is selected. For $\theta \neq 0$
the minimum of the fermion contribution moves
and as $\theta$ crosses $\pi/N_c$ the absolute minimum of the
potential moves to a different center element (RW transition).

In the case of an isospin chemical potential instead, 
the chemical potential is $\mu$ for half of the fermionic flavors 
and $-\mu$ for the other half. The effective 
potential therefore looks as follows:
\begin{eqnarray}
&& V_{\rm eff} = \frac{\pi^2T^4}{24}\sum_{j,k=1}^{N_c}
\left(1-\left(\left[\frac{\phi_{j}}{\pi}-\frac{\phi_{k}}{\pi}\right]_{\rm mod2}
-1\right)^{2}\right)^{2}
\nonumber\\
&& - \frac{\pi^2T^4}{12} \sum_{j=1}^{N_c} \left[ \sum_{i=1}^{N_f/2} 
\left(1-\left(\left[\frac{\phi_{j}
  + \theta}{\pi}+1\right]_{\rm mod2}-1\right)^{2}\right)^{2} \right. 
\nonumber \\
&& + \left. \sum_{i=N_f/2+1}^{N_f} \left(1-\left(\left[\frac{\phi_{j}
  - \theta}{\pi}+1\right]_{\rm mod2}-1\right)^{2}\right)^{2} \right] 
\label{effpotiso}
\end{eqnarray}

In this case the fermion contribution is left unchanged as 
$\phi_j \to - \phi_j$, i.e. it is symmetric under complex conjugation
of the Polyakov loop. However it easily verified that 
as $\theta$ crosses a critical value $\theta_c \simeq 0.5168\ \pi$,
the two degenerate minima corresponding to complex center elements becomes
deeper than then one corresponding to a real center element, hence
for  $\theta > \theta_c$ the conjugation symmetry of the Polyakov loop
is spontaneously broken and one of the two complex minima is selected.

Notice that the above effective potential is computed at one loop and
for zero mass fermions. Higher order and finite quark mass corrections
do not affect the location of the RW transition in the case of an
imaginary baryon chemical potential, which is fixed at 
$\theta = (2 k + 1) \pi/N_c$ by symmetry. The same corrections may 
however influence the location of $\theta_c$ for isospin chemical
potential, hence $\theta_c$ may be quark mass and temperature
dependent. Indeed, as shown in Fig.~\ref{su3_phase_diag}, numerical
simulations locate approximately $\theta_c \sim 0.48\ \pi$ as
the point where the unphysical RW-like line meets the analytic
continuation of the physical transition line. 

A more detailed 
analysis of the QCD phase diagram in the presence of imaginary 
baryon and isospin chemical potentials goes beyond our present
purposes and will be presented elsewhere~\cite{progress}.

\subsection{Results for the critical line at finite isospin}
\label{su3_results}

In Table~\ref{su3_isospin_data} we summarize our determinations 
of the critical couplings for the finite isospin SU(3) theory 
on a 8$^3\times 4$ lattice with fermionic mass $am$=0.1. 

\begin{table}[htbp]
\setlength{\tabcolsep}{0.5pc}
\centering
\caption[]{Summary of the values of $\beta_c(\mu^2)$ for SU(3) with finite
isospin density on the 8$^3\times 6$ lattice with fermionic mass $am$=0.1.} 
\begin{tabular}{dd}
\hline
\hline
\multicolumn{1}{c}{\hspace{0.70cm}$\mu^2/(\pi T)^2$} &
\multicolumn{1}{c}{\hspace{1cm}$\beta_c$} \\
\hline
-0.2304   & 4.842(1)     \\
-0.2209   & 4.834(1)     \\
-0.2116   & 4.8625(10)   \\
-0.198025 & 4.815(1)     \\
-0.1849   & 4.807(1)     \\
-0.16     & 4.7920(7)    \\
-0.126025 & 4.7710(7)    \\
-0.09     & 4.750(1)     \\
-0.0625   & 4.733(1)     \\
-0.04     & 4.7210(8)    \\
-0.01     & 4.703(1)     \\
 0.       & 4.697(1)     \\
 0.04     & 4.6715(15)   \\
 0.09     & 4.64125(100) \\
 0.16     & 4.5915(12)   \\
 0.2025   & 4.5585(10)   \\
\hline
\hline
\end{tabular}
\label{su3_isospin_data}
\end{table}

\begin{figure}[htbp]
\includegraphics*[height=0.295\textheight,width=0.95\columnwidth]
{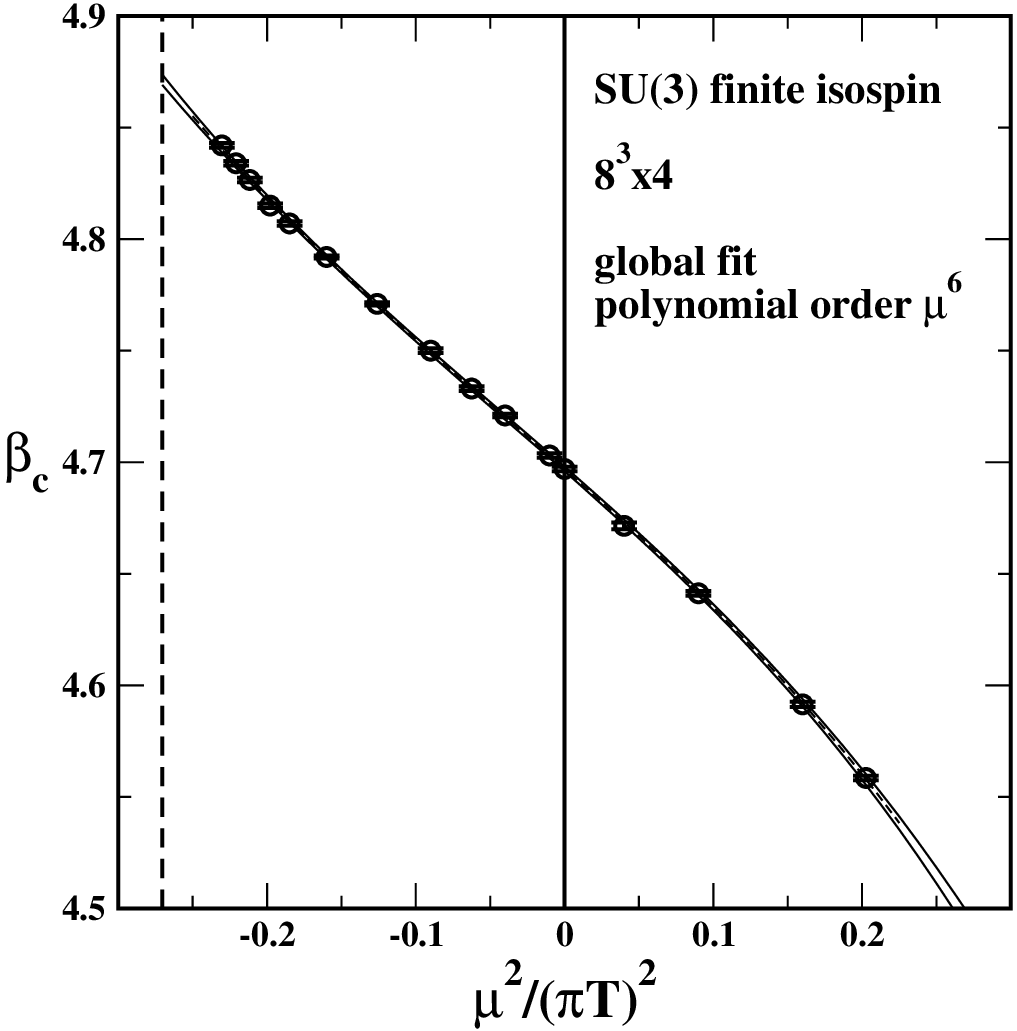}
\caption[]{Critical couplings obtained in SU(3) with finite isospin 
density on a 8$^3\times 4$ lattice with $am$=0.1, together with a polynomial 
fit of order $\mu^6$ to {\em all} data.}
\label{su3_crit_global}
\end{figure}

We observe from the very beginning that data for $\beta_c(\mu^2)$ for both 
$\mu^2\leq 0$ and $\mu^2>0$ can be globally fitted by an analytic function 
(a polynomial of third order in $\mu^2/(\pi T)^2$ nicely works). 
The fit parameters of 
the global fit are given in the last line of Table~\ref{su3_polin_fit}, 
while Fig.~\ref{su3_crit_global} shows how the fit compares with the data.

The question is if there are interpolations of the critical couplings at 
$\mu^2 \leq 0$ only, that, when continued to $\mu^2 > 0$, agree with the 
critical couplings directly determined in the latter region.

\begin{table*}[htbp]
\setlength{\tabcolsep}{0.1pc}
\centering
\caption[]{Parameters of the fits to the critical couplings 
in SU(3) with finite isospin density on the 8$^3\times 4$ lattice with 
fermionic mass $am$=0.1 according to the fit function
$\beta_c (\mu^2) = a_0 + a_1 \mu^2/(\pi T)^2 + a_2 \mu^4/(\pi T)^4 
+ a_3 \mu^6/(\pi T)^6 + a_4 \mu^8/(\pi T)^8 + a_5 \mu^{10}/(\pi T)^{10}$. 
Blank columns stand for terms not included
in the fit. The last column reports the largest value of $\mu^2/(\pi T)^2$
included in each fit. The $^*$ in the last but one line means ``constrained
parameter''.}
\begin{tabular}{ddddddcc}
\hline
\hline
\multicolumn{1}{c}{\hspace{1cm}$a_0$} & \multicolumn{1}{c}{\hspace{1cm}$a_1$} &
\multicolumn{1}{c}{\hspace{1cm}$a_2$} & \multicolumn{1}{c}{\hspace{1cm}$a_3$} &
\multicolumn{1}{c}{\hspace{1cm}$a_4$} & \multicolumn{1}{c}{\hspace{0.5cm}$a_5$} &
$\chi^2$/d.o.f.                       & $\mu_{\rm max}^2/(\pi T)^2$ \\
\hline
4.69535(79) & -0.6153(53) &           &           &          &           & 6.33 & 0. \\
4.6982(13)  & -0.529(26)  &  0.37(11) &           &          &           & 2.24 & 0. \\
4.6967(17)  & -0.636(72)  & -0.83(74) & -3.4(2.1) &          &           & 0.88 & 0. \\
4.6966(19)  & -0.638(63)  & -0.81(32) & -3.0(1.2) & 1.3(5.0) &           & 0.97 & 0. \\
4.6968(17)  & -0.621(24)  & -0.50(26) & -1.54(45) & 1.00(53) & -10.(19.) & 0.94 & 0. \\
4.6974(8)   & -0.596(12)^*& -0.43(14) & -2.36(61) &          &           & 1.07 & 0. \\
4.6970(10)  & -0.585(12)  & -0.190(44)& -1.528(35)&          &           & 1.10 & 0.2025 \\
\hline
\hline
\end{tabular}
\label{su3_polin_fit}
\end{table*}

\begin{figure*}[htbp]
\includegraphics*[height=0.295\textheight,width=0.95\columnwidth]
{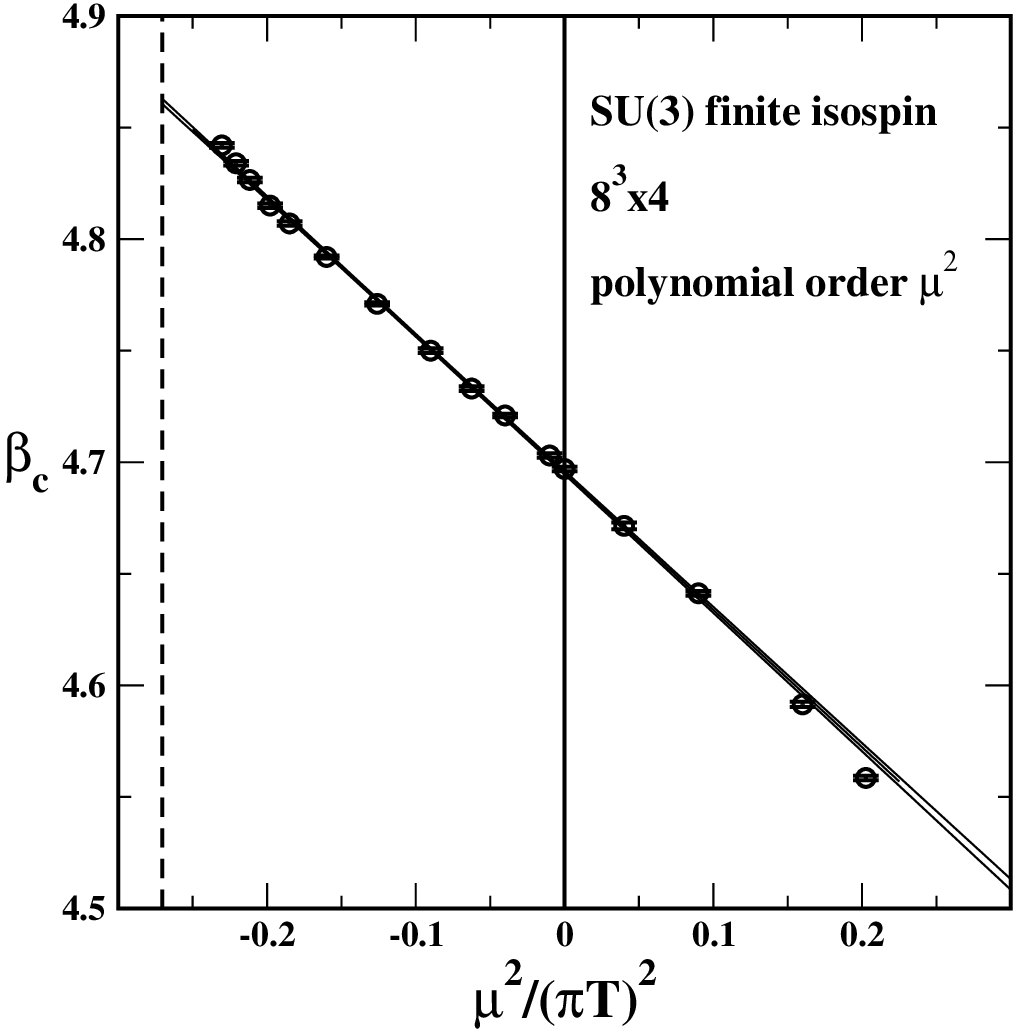}
\includegraphics*[height=0.295\textheight,width=0.95\columnwidth]
{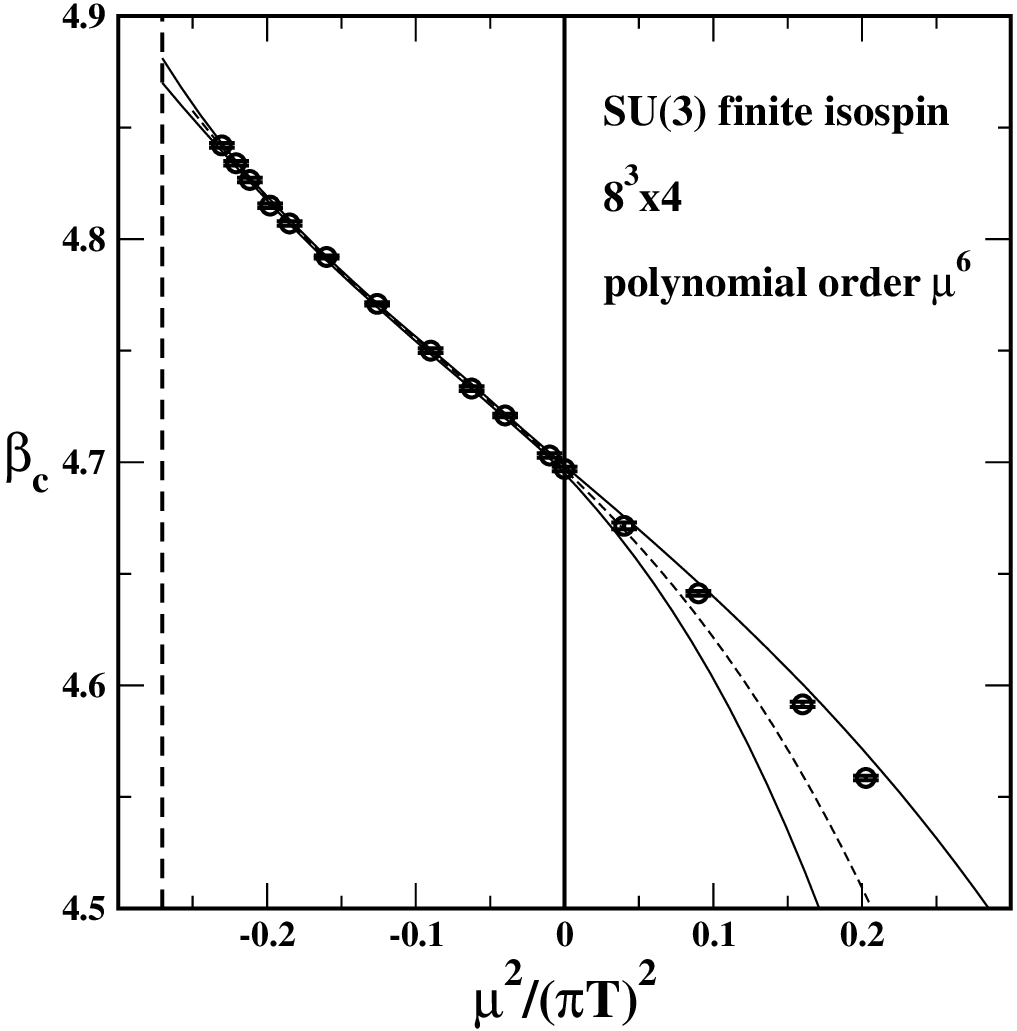}
\caption{Critical couplings obtained in SU(3) with finite isospin density 
on a 8$^3\times 4$ lattice with $am$=0.1, together with a polynomial fit of
order $\mu^2$ (left) and $\mu^{6}$ (right) to data with $\mu^2 \leq 0$.}
\label{su3_crit_polin}
\end{figure*}

We have tried several kind of interpolations of the critical couplings 
at $\mu^2 \leq 0$. 
At first, we have considered interpolations with polynomials up to
order $\mu^{10}$ (see Table~\ref{su3_polin_fit} for a summary of the
resulting fit parameters). We can see that data at $\mu^2 \leq 0$ are 
precise enough to be sensitive to terms beyond the order $\mu^2$; indeed,
a good $\chi^2$/d.o.f. is not achieved before including terms up to the 
order $\mu^6$, in agreement with the outcome of the global fit discussed
above. The extrapolation to $\mu^2 > 0$ for the polynomial of order 
$\mu^6$ is shown in Fig.~\ref{su3_crit_polin} (right), in comparison
with that obtained from the polynomial of order $\mu^2$ 
(Fig.~\ref{su3_crit_polin} (left)). The extrapolation agrees with direct 
determinations of $\beta_c(\mu^2)$, within the 95\% CL band, only if 
nonlinear terms in $\mu^2$, at least up to $\mu^6$, are taken into account.

\begin{table*}[htbp]
\setlength{\tabcolsep}{-0.15pc}
\centering
\caption[]{Parameters of the fits to the critical couplings 
in SU(3) with finite isospin density on the 8$^3\times 4$ lattice with 
fermionic mass $am$=0.1 according to the fit function
$\beta_c (\mu^2) = (a_0 + a_1 \mu^2/(\pi T)^2 + a_2 \mu^4/(\pi T)^4 
+ a_3 \mu^6/(\pi T)^6)/(1 + a_4 \mu^2/(\pi T)^2 + a_5 \mu^4/(\pi T)^4 
+ a_6 \mu^6/(\pi T)^6)$. Blank columns stand for terms not included in the 
fit.}
\begin{tabular}{dddddddc}
\hline
\hline
\multicolumn{1}{c}{\hspace{1cm}$a_0$} & \multicolumn{1}{c}{\hspace{1cm}$a_1$} &
\multicolumn{1}{c}{\hspace{1cm}$a_2$} & \multicolumn{1}{c}{\hspace{1cm}$a_3$} &
\multicolumn{1}{c}{\hspace{1cm}$a_4$} & \multicolumn{1}{c}{\hspace{1cm}$a_5$} &
\multicolumn{1}{c}{\hspace{1cm}$a_6$} & $\chi^2$/d.o.f. \\
\hline
4.6982(13)  &  2.21(76)   &             &           & 0.58(16)   &              &           & 2.04 \\
4.69748(79) & 18.3737(83) &             &           & 4.0328(17) &   0.4711(58) &           & 0.50 \\
4.6970(12)  & 15.034(17)  &             &           & 3.3287(35) &   0.461(17)  & 0.236(55) & 0.49 \\
4.69730(85) & 17.3130(99) &  -2.177(33) &           & 3.8089(21) &              &           & 0.44 \\
4.6967(14)  &  2.124(34)  & -20.38(17)  &           & 0.5855(73) &  -4.136(35)  &           & 0.74 \\
4.6971(13)  &  1.920(18)  & -58.143(93) &           & 0.5353(38) & -12.277(19)  &-1.332(67) & 0.55 \\
4.6971(11)  & 16.284(13)  &  -2.202(62) & -0.46(20) & 3.5928(27) &              &           & 0.46 \\
4.6971(13)  &  4.503(19)  & -44.770(95) &  4.82(33) & 1.0851(39) &  -9.366(20)  &           & 0.54 \\
4.6971(14)  &  4.956(19)  & -42.960(97) &  4.55(40) & 1.1816(39) &  -8.969(20)  &-0.007(82) & 0.65 \\
\hline
\hline
\end{tabular}
\label{su3_ratio_fit}
\end{table*}

\begin{figure*}[htbp]
\includegraphics*[height=0.295\textheight,width=0.95\columnwidth]
{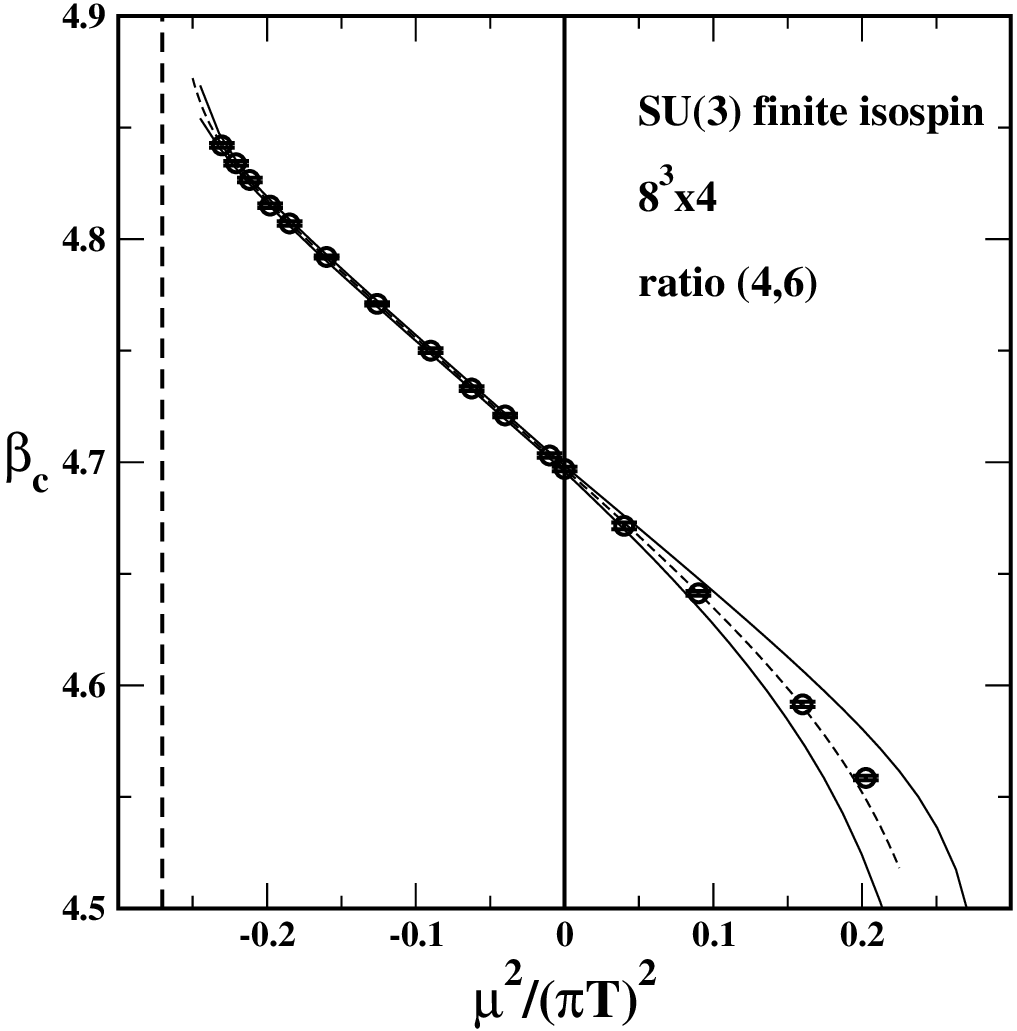}
\includegraphics*[height=0.295\textheight,width=0.95\columnwidth]
{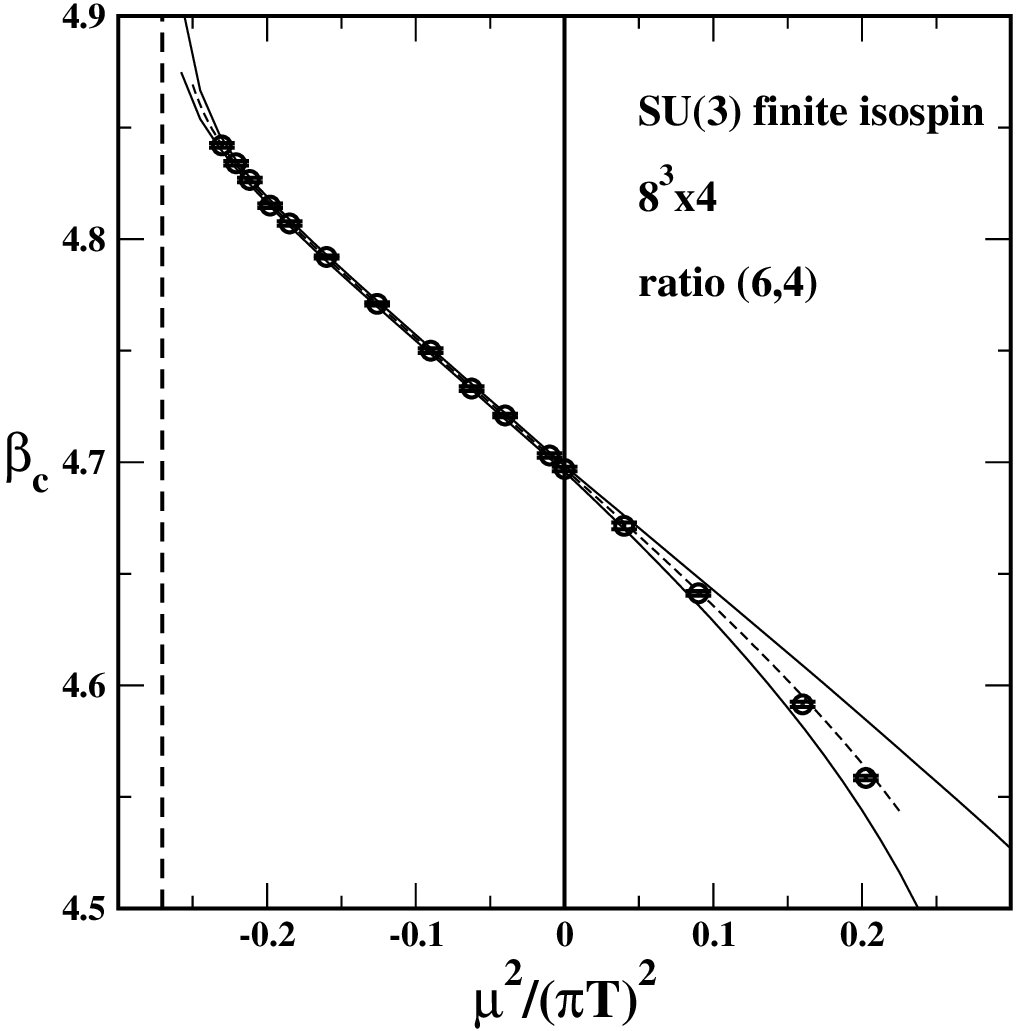}
\caption{Same as Fig.~\ref{su3_crit_polin}, with the fitting function given by 
the ratio of a 4th to 6th order polynomial (left) and the ratio of a 6th to 
4th order polynomial (right).}
\label{su3_crit_ratio}
\end{figure*}

Then, we have considered interpolations with ratios of polynomials of
order up to $\mu^{6}$ (see Table~\ref{su3_ratio_fit} for a summary of the
resulting fit parameters). In all but one case we got good fits to the
data at $\mu^2 \leq 0$, but only two extrapolations to $\mu^2 > 0$ compare
well with numerical data in that region: the ratio of a 4th to 6th order 
polynomial and the ratio of a 6th to 4th order polynomial (see 
Fig.~\ref{su3_crit_ratio}). It is interesting to observe that the two 
interpolations which ``work'' have in common the number of parameters.

\begin{table}[htbp]
\setlength{\tabcolsep}{0.3pc}
\centering
\caption[]{Values of the parameter $a_1$ resulting from the fit to the 
critical couplings in SU(3) with finite isospin density on the 8$^3\times 6$ 
lattice with fermionic mass $am$=0.1 according to the fit function
$\beta_c (\mu^2) = a_0 + a_1 \mu^2/(\pi T)^2$, in the interval 
$[\mu^2_{\rm min}/(\pi T)^2,0]$.}
\begin{tabular}{ddc}
\hline
\hline
\multicolumn{1}{c}{\hspace{0.5cm}$\mu_{\rm min}^2/(\pi T)^2$} & 
\multicolumn{1}{c}{\hspace{1cm}$a_1$} & $\chi^2$/d.o.f. \\
\hline
-0.04     & -0.60(5)    & -    \\
-0.0625   & -0.58(3)    & 0.39 \\
-0.09     & -0.585(21)  & 0.29 \\
-0.126025 & -0.586(12)  & 0.22 \\
-0.16     & -0.5915(88) & 0.39 \\
-0.1849   & -0.5931(77) & 0.39 \\
-0.198025 & -0.5945(69) & 0.39 \\
-0.2116   & -0.6010(62) & 1.83 \\
-0.2209   & -0.6078(57) & 3.72 \\
\hline
\hline
\end{tabular}
\label{su3_quad_fit}
\end{table}

\begin{figure}[htbp]
\includegraphics*[height=0.295\textheight,width=0.95\columnwidth]
{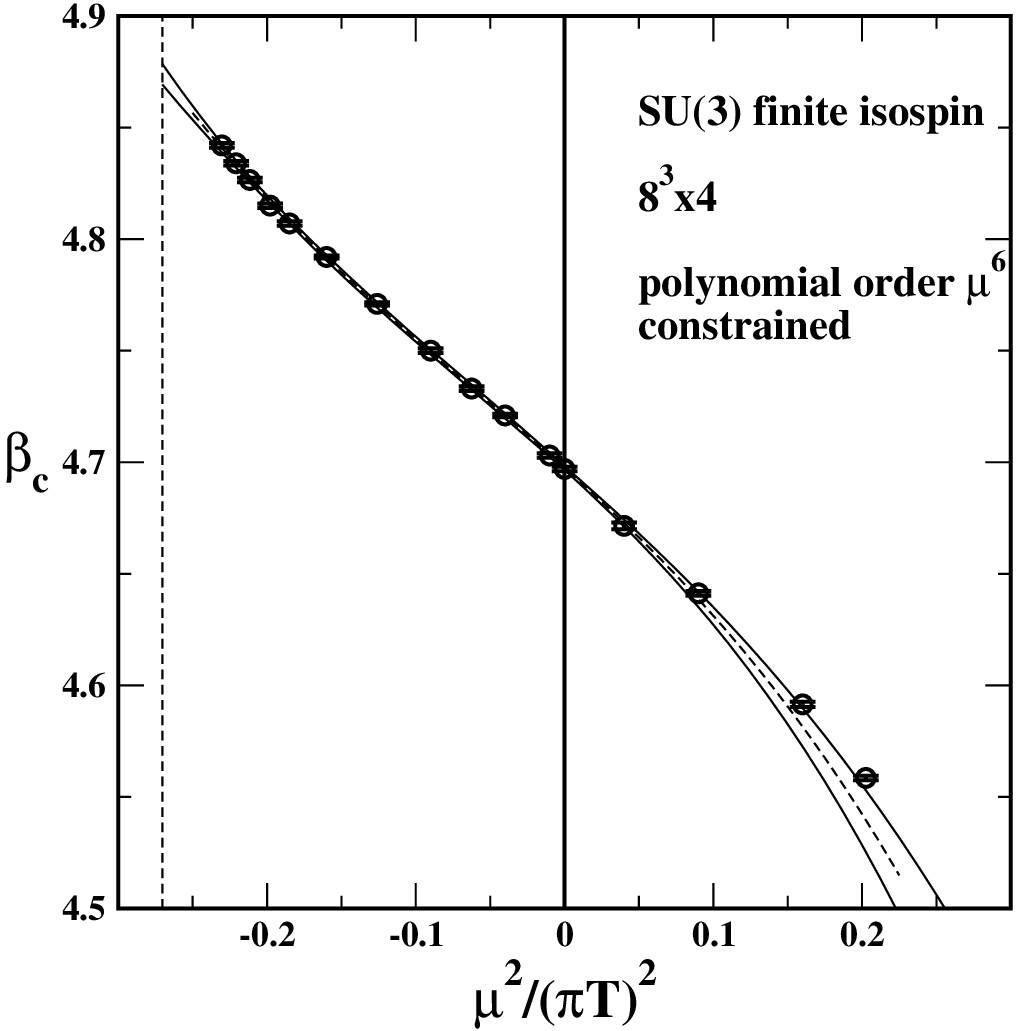}
\caption{Same as Fig.~\ref{su3_crit_polin}, with the fitting function given by 
a polynomial of order $\mu^6$ with the quadratic term constrained as explained in the text.}
\label{su3_crit_constrained}
\end{figure}

Both kinds of fits considered so far have evidenced that the role of
terms of order larger than $\mu^2$ cannot be neglected. Since the data
more sensitive to these terms are those farther from $\mu^2=0$, while data
closer to $\mu^2=0$ should be enough to fix the coefficient of the $\mu^2$ 
term in a polynomial interpolation, we followed a different strategy:
we performed a fit with a polynomial of the form $a_0+a_1 \mu^2/(\pi T)^2$
in the region $\mu^2_{\rm min}/(\pi T)^2 \leq \mu^2/(\pi T)^2 \leq 0$
and varied $\mu^2_{\rm min}$ in order to find the largest possible interval
in which the fit gives both a good $\chi^2$/d.o.f. and a stable value
(within errors) for the linear coefficient $a_1$; this fixes the value of
the parameter $a_1$ with some accuracy (see Table~\ref{su3_quad_fit}) 
\footnote{A similar strategy has been followed also in 
Ref.~\cite{deForcrand:2008vr}.}.
After that, we fixed $a_1$ at 0.590 (allowing an arbitrary variability of 
0.006) and repeated the fit on all available data at $\mu^2 \leq 0$
with a polynomial of the form $a_0+a_1 \mu^2/(\pi T)^2+a_2 \mu^4/(\pi T)^4
+a_3 \mu^6/(\pi T)^6$. The resulting interpolation and its extrapolation
to the region $\mu^2>0$ are shown in Fig.~\ref{su3_crit_constrained}.
The comparison with critical couplings at $\mu^2>0$ is good and the 95\% CL
band is narrower than in the unconstrained $\mu^6$-polynomial fit 
(see Fig.~\ref{su3_crit_polin} (right)), meaning that this procedure
leads to increased predictivity for the method of analytic continuation.

\begin{figure*}[htbp]
\includegraphics*[height=0.295\textheight,width=0.95\columnwidth]
{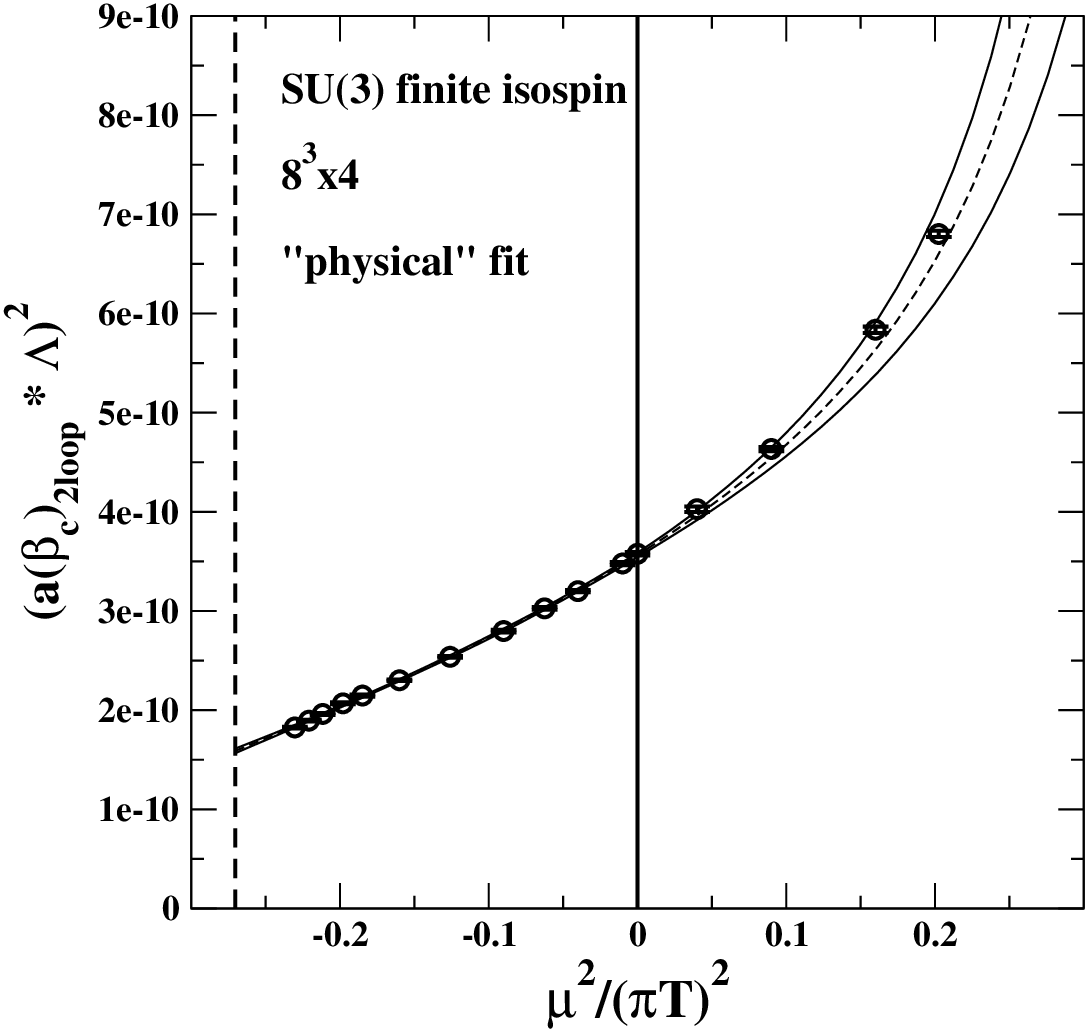}
\includegraphics*[height=0.295\textheight,width=0.95\columnwidth]
{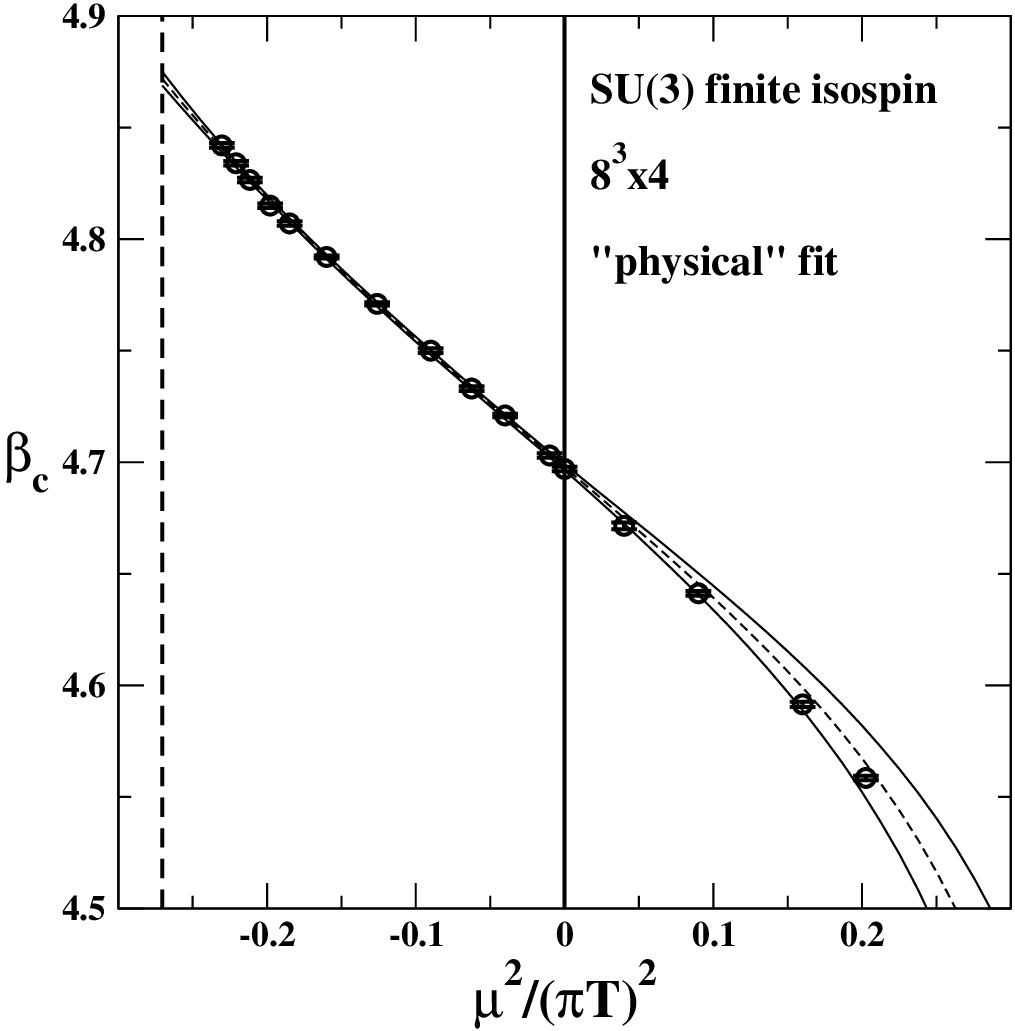}
\caption{Values of $[a(\beta_c(\mu^2))\Lambda]^2$ (left) 
and $\beta_c(\mu^2)$ (right) in SU(3) with finite isospin 
density on a 8$^3\times 4$ lattice with $am$=0.1, together with the 
fit to data with $\mu^2\leq 0$ according to the fit 
function~(\ref{phys_fit}).} 
\label{su3_phys_fit}
\end{figure*}

\begin{figure*}[htbp]
\includegraphics*[height=0.295\textheight,width=0.95\columnwidth]
{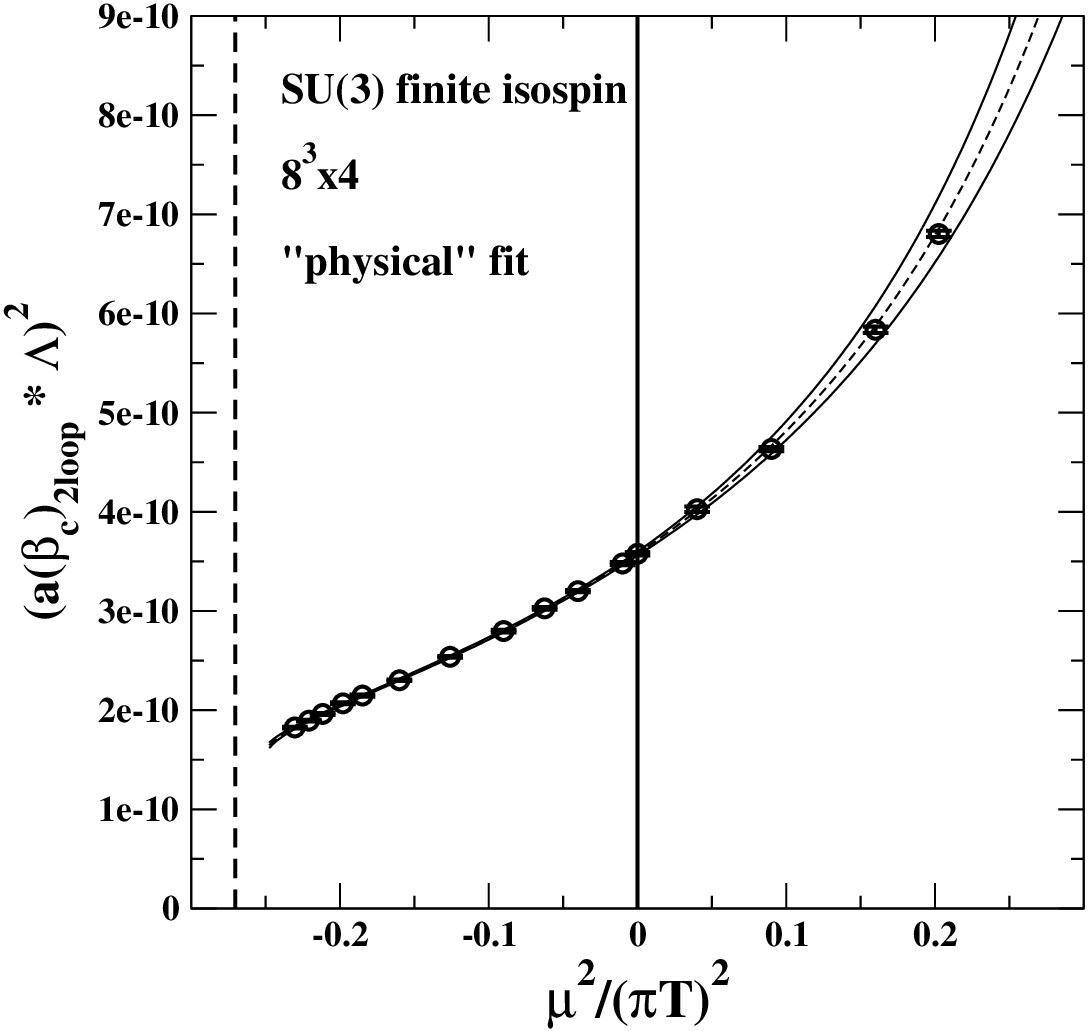}
\includegraphics*[height=0.295\textheight,width=0.95\columnwidth]
{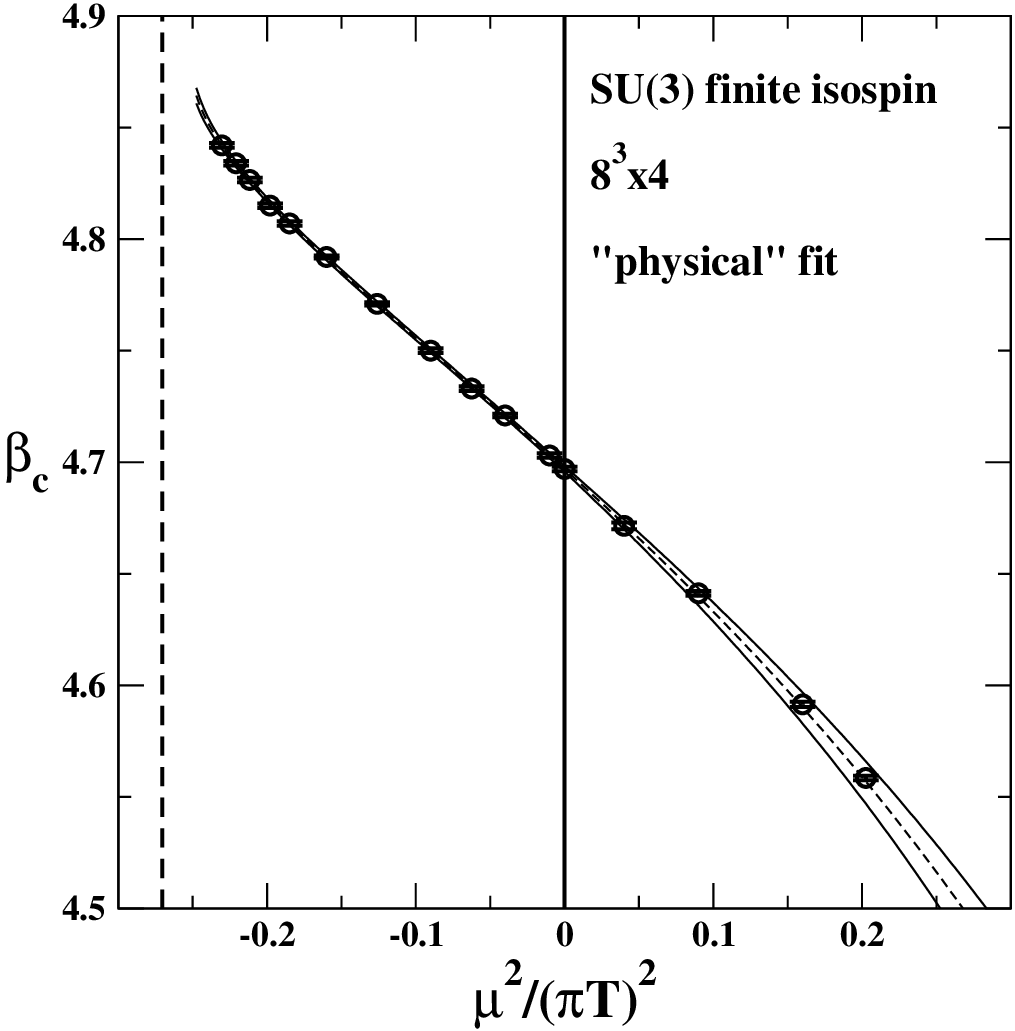}
\caption{Same as Fig.~\ref{su3_phys_fit} with the fit to data with 
$\mu^2\leq 0$ according to the fit function~(\ref{phys_crit_new}).} 
\label{su3_phys_fit_new}
\end{figure*}

At last, we have attempted a fit strategy never used so far in our studies
on the present subject and, to our knowledge, nowhere in the literature
on the analytic continuation from imaginary chemical potential. The idea
is to write the interpolating function in {\em physical units} and to
deduce from it the functional dependence of $\beta_c$ on $\mu^2$, after
establishing a suitable correspondence between physical and lattice units.
The natural, dimensionless variables of our theory are $T/T_c(0)$,
where $T_c(0)$ is the critical temperature at zero chemical
potential, and $\mu/(\pi T)$. The question that we want to answer is 
if fitting directly the dependence of $T/T_c(0)$ on $\mu/(\pi T)$  
may lead to increased predictivity 
for analytic continuation. We shall name this kind of fits, made directly
in terms of the dimensionless physical variables of the theory, 
as ``physical'' fits.

While $\mu/(\pi T)$ is one of the dimensionless variables used in our 
simulations, $T/T_c(0)$ is not and must be deduced from the relation
$T=1/(N_t a(\beta))$, 
where $N_t$ is the number of lattice sites in the temporal direction
and $a(\beta)$ is the lattice spacing at a given $\beta$~\footnote{
Strictly speaking the lattice spacing depends also on the bare quark 
mass, which in our runs slightly changes as we change $\beta$ since we fix $a m$.
However in the following evaluation, which is only based on the perturbative
2-loop $\beta$-function, we shall neglect such dependence.
}. Since our 
determinations for $\beta_c$ range between $\simeq$ 4.5585 and 
$\simeq$ 4.842 (see Table~\ref{su3_isospin_data}) it can make sense to use  
for $a(\beta)$ the perturbative 2-loop expression with $N_c=3$ and $N_f=8$.

We have tried several different fitting functions and report two cases
which work particularly well. The first is given by the following
3-parameter function:
\beq
\left[\frac{T_c(\mu)}{T_c(0)}\right]^2=\frac{1+B\mu^2/(\pi T_c(\mu))^2}
{1+A\mu^4/(\pi T_c(\mu))^4}\; 
\label{phys_crit}
\eeq
leading to the following implicit relation between $\beta_c$ and 
$\mu^2$:
\begin{eqnarray}
a(\beta_c(\mu^2))^2|_{\text{ 2-loop}} &=& a(\beta_c(0))^2|_{\text{2-loop}}
\nonumber \\
&\times& \frac{1+A\mu^4/(\pi T_c(\mu))^4}{1+B\mu^2/(\pi T_c(\mu))^2}\;.
\label{phys_fit}
\end{eqnarray}
In Fig.~\ref{su3_phys_fit}(left) we compare the values of 
$a(\beta_c(\mu^2))^2|_{|\rm 2-loop}$ obtained using our determinations for
$\beta_c(\mu^2)$ with the interpolation according to the fit 
function~(\ref{phys_fit})
to data in the region $\mu^2\leq 0$: one can see that the extrapolation to the
region $\mu^2>0$ behaves very well. In Figs.~\ref{su3_phys_fit}(right) we 
have rescaled the vertical axis in order to give the behavior of 
$\beta_c(\mu^2)$ versus the isospin chemical potential.
The values of the fit parameters are
\begin{eqnarray}
\beta_c(0) = 4.6977(13) \;,\;\;\;\;\;\; A &=& -3.25(26) \;, \nonumber \\
B &=& -2.62(12)\;,
\end{eqnarray}
with $\chi^2$/d.o.f.=1.33. 
We observe that $\beta_c(0)$ is in nice agreement with the direct 
determination (see Table~\ref{su3_isospin_data}).

As an alternative function for the shape of the critical line, we have 
also tried the following
\beq
\frac{T_c(\mu)}{T_c(0)}=
\left\{
\begin{array}{ll}
A+(1-A)\left[\cos\left(\frac{\mu}{T}\right)\right]^B\;, & \mu^2\leq 0 \\
A+(1-A)\left[\cosh\left(\frac{\mu}{T}\right)\right]^B\;, 
& \mu^2 > 0 \;, \\
\end{array}
\right.
\label{phys_crit_new}
\eeq
which explicitly encodes the expected periodicity of the partition
function for imaginary $\mu$.  
The fit to data at imaginary $\mu$ is very good and its extrapolation
to the real chemical potential side compares impressively well with data
(see Fig.~\ref{su3_phys_fit_new}). The resulting fit parameters are
\begin{eqnarray}
\beta_c(0) = 4.6969(12) \;,\;\;\;\;\;\; A &=& 1.508(15) \;, \nonumber \\
B &=& 0.560(32)\;,
\end{eqnarray}
with $\chi^2$/d.o.f.=0.39. This function is a good candidate to 
parametrize the critical line for small values of $\mu/T$.

In both cases, Eqs.~(\ref{phys_fit}) and~ \ref{phys_crit_new}, the 
physical fit has worked very well and with a reduced number of parameters
with respect to our previous fits, leading to increased predictivity and 
consistency with data at real chemical potentials. In both cases one can easily 
check that the adopted functions are not appropriate for a continuation 
of the critical line down to the $T = 0$ axis, but this is not the aim 
of our study since such extrapolation would be questionable anyway. 

We have used other functional forms for $T_c(\mu)$, inspired by the
Gross-Neveu model~\cite{GN} and by the random matrix theory~\cite{RM},
and found that they allow one to interpolate data at $\mu^2\leq 0$, but their
extrapolation does not match data at $\mu^2 >0$. 

Finally, in order to compare the efficiency of the different predictions
of the critical line, we present in Table~\ref{su3_compare_0.2} the extrapolated 
value of $\beta_c$ at a common value of the real chemical potential
($\mu^2/(\pi T)^2=0.2$) for all
the interpolation methods adopted. We can see that the last two methods
listed (constrained and physical fit) have a narrower error band,
the physical fit having a better agreement with data.

\begin{table}[htbp]
\setlength{\tabcolsep}{1.pc}
\centering
\caption[]{Extrapolated values of $\beta_c (\mu^2)$ at $\mu^2/(\pi T)^2=0.2$
for the interpolation methods adopted in 
Figs.~\ref{su3_crit_polin}-\ref{su3_phys_fit_new}.} 
\begin{tabular}{dl}
\hline
\hline
\multicolumn{1}{c}{\hspace{0.5cm}$\beta_c$} & kind of interpolation \\
\hline
4.51(6)              & Fig.~\ref{su3_crit_polin}(right) \\
4.55(3)              & Fig.~\ref{su3_crit_ratio}(left)  \\
4.56^{+0.03}_{-0.02} & Fig.~\ref{su3_crit_ratio}(right) \\
4.54^{+0.02}_{-0.01} & Fig.~\ref{su3_crit_constrained}  \\
4.57^{+0.01}_{-0.02} & Fig.~\ref{su3_phys_fit} (right)  \\
4.56(1)              & Fig.~\ref{su3_phys_fit_new} (right)\\
\hline
\hline
\end{tabular}
\label{su3_compare_0.2}
\end{table}

\section{Discussion}

The present study is part of a larger project which, based
on the investigation of QCD-like theories which are free of the sign
problem, is aimed at testing the validity of the method of analytic continuation
and at improving its predictivity, in view of its application to real QCD.

In particular, we have presented results concerning 
the analytic continuation of the critical line 
in 2-color QCD and in QCD with a finite density of isospin charge.
We have detected some common features and developed some strategies,
which could apply and be useful also for real QCD at finite baryon density. Let 
us briefly summarize them.

\begin{enumerate}

\item[(i)] Non-linear terms in the dependence of the pseudocritical 
coupling $\beta_c$ on $\mu^2$ in general cannot be neglected.
A polynomial of order $\mu^6$ seems
to be sufficient in all explored cases. The prediction for the pseudocritical 
couplings at real chemical potentials may be wrong if data 
at imaginary $\mu$ are fitted according to a linear dependence. 

\item[(ii)] The coefficients of the linear and non-linear terms in $\mu^2$ in
a Taylor expansion of $\beta_c(\mu^2)$ are all negative. That often implies
subtle cancellations of nonlinear terms at imaginary chemical 
potentials ($\mu^2 < 0$) in the region available for analytic
continuation (first RW sector). The detection of such terms,
from simulations at $\mu^2 < 0$ only, may be difficult and requires an
extremely high accuracy.
As a matter of fact, the simple use of a sixth order polynomial to fit 
data at imaginary $\mu$ leads to poor predictivity, which is slightly 
improved if ratio of polynomials are used instead.

\item[(ii)] An increased predictivity is achieved if the following strategy is 
adopted: first, the linear term in $\mu^2$ is fixed from data at small values
of $\mu^2$ only. Then a nonlinear fit (e.g. with a sixth order
polynomial) is performed, keeping the $\mu^2$ term constrained. 

\item[(iii)] We have proposed a new, alternative ansatz to parametrize the
critical line directly in physical units in the $T,\mu$ plane 
(instead than in the $\beta,\mu$ plane) and given two explicit realizations.
We have shown that this ``physical'' ansatz provides a very good description 
of the critical line, moreover with a reduced number of parameters, 
and leads to an increased predictivity, comparable to that
achieved by the ``constrained'' fit. 

\end{enumerate}

We plan to apply our experience to the determination of 
the critical line for real QCD in the near future.
The lesson from the present study is that the preferred
extrapolation methods to be considered in that case 
are a nonlinear polynomial, at least of sixth order 
in $\mu$, in which the $\mu^2$ term is constrained, or rather the new 
``physical'' ansatz for the critical line that we have proposed.
The two methods have shown comparable predictivity.
The former requires dedicated simulations at small
chemical potentials, in order to better fix the $\mu^2$ term,
which could also be fixed by using alternative methods like
reweighting or Taylor expansion. The latter is more appealing
since it describes the critical line with a reduced number 
of parameters: this is very likely at the origin of
its increased predictivity and is probably related
to the fact that one is looking for a direct relation between
two physical quantities ($T/T_c(0)$ and $\mu/T$) in this
case, without going through bare lattice quantities.

\section{Acknowledgments}

We acknowledge the use of the computer facilities at the INFN apeNEXT 
Computing Center in Rome and of the PC clusters of CNAF-Bologna and INFN-Bari.
We thank Ph.~de~Forcrand and M.P.~Lombardo for very useful discussions.

\end{document}